# Is there a size premium for nations?[1]

Jože P. Damijan[2]

Sandra Damijan[3]

Osiris Jorge Parcero[4]

## Abstract

This paper examines whether there is a premium in country size. We study whether there are significant gains from being a small or a large country in terms of certain socioeconomic indicators, and how large this premium is. Using panel data for 200 countries over 50 years, we estimate premia for various sizes of nations across a variety of key economic and socio-economic performance indicators. We find that smaller countries are richer, have larger governments, and are more prudent in terms of fiscal policies than larger ones. Smaller countries seem to be subject to higher absolute and per capita costs for provision of essential public goods, which may lower their socio-economic performance in terms of health and education. In terms of economic performance, small countries seem to do better than large countries, compensating for smallness by relying on foreign trade and foreign direct investment. The latter comes at the cost of higher vulnerability to external shocks, resulting in higher volatility of growth rates.

This paper's findings offer essential guidance to policymakers, international organizations, and business researchers in general, especially those assessing a country's economic or socio-economic performance or potential. The study implies that comparisons with medium-sized or large countries may be of little utility in predicting the performance of small countries.

---

[1] Financial support of VIVES Institute at the University Leuven is greatly acknowledged. We thank Abdul Rashid, Almat Kenen, Yerkezhan Kenzheali, and Zhibek Kassymkanova for excellent help with data collection and processing, and Annelore Van Hecke, Joep Konings and Črt Kostevc for providing valuable comments to an earlier draft of the paper.

[2] Faculty of Economics, University of Ljubljana, Kardeljeva pl. 17, 1000, and University of Leuven (joze.damijan@ef.uni-lj.si).

[3] Faculty of Economics, University of Ljubljana, Kardeljeva pl. 17, 1000 (sandra.damijan@ef.uni-lj.si).

[4] International School of Economics, Kazakh-British Technical University, 59 Tole Bi Street, Almaty, Kazakhstan. 050000 (osirisjorge.parcero@gmail.com). ORCID 0000-0002-6899-7068

# 1. Introduction

Having economies of scale is among the key advantages characterizing large countries, as such economies increase growth and welfare. New trade and economic geography theories include agglomeration effects suggesting that economic activity will more likely concentrate close to large markets (Krugman, 1991; Krugman and Venables, 1995; Fujita et al., 1999). Similarly, scale effects are generic to endogenous growth models, implying that an economy's size positively affects a country's long-term growth rate (Aghion and Howitt, 1998). The most prominent scale effects, however, are nested in public economics. Alesina and Spolaore (1997, 2003) list five benefits of large population size: (i) lower per-capita costs of public goods; (ii) cheaper per-capita defense and military costs; (iii) greater productivity due to specialization; (iv) greater ability to provide regional insurance; and (v) greater ability to redistribute income within borders.

According to IMD World Competitiveness Center (2012), there are only three large countries among the top ten most competitive countries in 2012, and only five among the top 20. Thus, in contrast to predictions based on prevailing theory, smallness does not seem to translate into economic underperformance. Several factors make small countries flexible in terms of governance. The most important contributor to the economic success of small nations seems to involve their lower degree of cultural and political diversity, which may lead to greater ability to reach political consensus on vital economic policy issues. Small countries can insure against some disadvantages of smallness through membership in supranational organizations and trade blocs. They overcome the handicap of having small national markets by creating extensive trade links, thus reaping scale effects in at least some sectors.

Empirical analysis thus far has failed to produce unambiguous evidence for the notion that national size is a factor – either positive or negative – in economic performance. Numerous studies in Robinson's 1960 compendium (Robinson, 1960) tested for the impact of economies of scale on country performance, finding that having economies of scale or not is for the most part unimportant. These findings were later confirmed using more recent data by Damijan (1996) and Salvatore et al. (2001). Barro and Sala-i-Martin (1995) provide limited evidence of scale effects on growth. Alesina et al., (2005) test whether the effect of size on growth depends on country openness, finding only moderately supportive evidence. Rose (2006) examines the impact of size on many country characteristics. He finds that small countries are richer and more open to international trade than large ones but are not systematically different otherwise.

This paper aims to identify a size premium for nations in terms of certain socio-economic indicators. In particular, we investigate the impact of country size on levels of income, long-term economic growth, volatility of growth, openness to trade and foreign direct investment (FDI), government budget and current account balances, size of government and public debt, inflation, standards of living, income distribution, health, education, infrastructure development, levels of democracy, and several other socio-economic indicators. Our empirical investigation uses a comprehensive database of more than 200 countries between 1960 and 2010.

We find that small countries do behave differently from large ones, even after we control for several country-specific characteristics. Country size matters in many ways, though sometimes the effect of smallness is positive and sometimes negative. We find that small countries are richer and have larger governments but are also more prudent in fiscal policies and run smaller public debts than large ones. Small countries seem to pay higher per capita costs for provision of essential public goods and seem to get less for their penny in terms of performance in health and education. This last does not hold for military spending in small countries, which have both lower spending and lower tendency to engage in armed conflict than large countries. Smallness does not result in higher income inequality or in less democracy. To a large extent, small countries compensate for the smallness of their domestic markets by relying on foreign trade and FDI. This reliance, however, increases their vulnerability to external shocks and results in higher growth volatility.

The following section briefly reviews the literature and posits a set of testable hypotheses. Section 3 proposes our empirical approach and introduces data. Section 4 presents and discusses results. The last section concludes.

# 2. Literature review and hypotheses

## 2.1. Overview of literature

Large countries may benefit from greater productivity than small ones due to specialization and competition effects associated with large domestic markets. This potential benefit of size was noted as far back as Adam Smith, who acknowledged that the extent of a market puts a limit on the advantages of division of labor. The notion that large market size is advantageous is crucial to three recent strains in literature: new trade theory, new economic geography, and endogenous growth models.

The concept of increasing returns to scale as a consequence of fixed costs in the production of differentiated goods plays a crucial role in new trade theory pioneered by Krugman (1979 and 1980) as well as by Ethier (1982) and Helpman (1984). In the context of this approach, the advantage of larger markets is that they can host more firms and hence allow for more goods varieties. New trade models posit that small countries can compensate for their smallness by becoming more open to trade. Empirical evidence supports this prediction; for instance, see Alesina and Wacziarg (1998) and Alesina (2003).

Scale effects are fundamental to endogenous growth models (Aghion and Howitt, 1998). Some of the ideas motivating this approach go back to Myrdal and Kaldor, who put forth the mechanisms of cumulative causation in the context of development and growth dynamics (Fujita, 2007; O'hara, 2008), but Romer (1986, 1987, 1990) and Lucas (1988) pioneered formal models. The primary sources of increasing returns in Romer (1987) are specialization and product differentiation, so the growth rate is directly proportional to the stock of human capital in an economy. Romer (1990) incorporates research spillovers, also positively affected by an economy's size. Moreover, in the three classes of endogenous growth models discussed by Jones (1999), there are what the author terms 'strong scale effects,' i.e., "the size of the economy affects either the long-run growth rate or the long-run level of per capita income." Several papers have stressed that larger market size enhances growth by raising the intensity of product market competition (Aghion and Howitt, 1998).

New economic geography models are also built on the concept of increasing returns to scale, but they go one step further than new trade theory, providing a unified theory of trade and production geographical localization. Moreover, as stated by Westlund (2020), "…while the endogenous growth theory, with its stress of human capital, ideas, and knowledge, rather explicitly is a theory for the knowledge economy, the new economic geography relies more on historic initial advantages …" It is thus no coincidence that Krugman (1991), in the closing commentary of his Economy and Trade, indicated that his message was somewhat a repetition of well-known ideas already present in the thinking of Marshall, Young, Myrdal, Hirschman, Pred, and Kaldor (Westlund, 2020).

Two influential papers are Helpman and Krugman (1987) and Krugman (1991), while Ottaviano and Puga (1998) provide an early survey of theoretical papers. One crucial implication of this literature is that the proportion of manufactures in larger countries is more prominent than would be predicted by their proportion in terms of population. In addition, a larger country is expected to enjoy higher real wages and welfare if trade costs are not close enough to zero. Overman et al. (2003), Redding (2010), and Damijan and Kostevc (2012) offer reviews of the empirical literature. Finally, it is important to stress that the new economic geography has influenced endogenous growth theory and the latter has slowly but certainly incorporated spatial factors (Bond-Smith and McCann, 2021).

An additional strand of literature is important for our work. For the most part, it also backs the claim that country size matters in terms of potential scale effects of public goods provision. Alesina and Spolaore (1997, 2003) and Alesina et al., (2005) recognize that four out of five of the benefits of a large population are nested in the public and political economics fields.

First, Alesina and Spolaore (1997, 2003) stress that the fixed cost of producing certain public goods leads to higher per-capita costs in smaller countries. This is particularly the case for monetary and financial systems, judicial systems, infrastructure for communications, police and crime prevention, and public health. Indeed, Alesina and Wacziarg (1998) show that government spending as a share of GDP decreases with population – i.e., larger countries have smaller per capita governments. Second, larger countries can realize cheaper per-capita defense and military costs and reduced probability of suffering foreign aggression (Latzko, 1993). To some extent, small countries may reduce the necessity of building robust defense capacity by entering into military alliances but also because they are less likely to engage in armed conflict (Hegre et al., 2013).

Third, larger countries can more easily benefit from centralizing the provision of certain public goods that involve strong externalities. In particular, larger countries may be more able to provide insurance to regions affected by adverse shocks (e.g., natural catastrophes). Forth, larger countries have a greater ability to redistribute income from more affluent regions to more impoverished ones. However, this could have the effect of encouraging wealthier regions to secede, resulting in partition of large countries into smaller ones. Thus, smallness may become endogenous to richness. In other words, in equilibrium, small countries will be those that can afford to maintain the costs of smallness, and hence, will tend to be richer. These arguments resemble well the notion by Alesina (2003) that country size is endogeneous to politico-economic forces, i.e. evolution of the size of countries is endogenous in a long run and driven by economic success and political reasons.

Several studies hypothesize that larger countries will have a higher tendency toward redistributive policies at the individual level. For instance, based on this strand of political science literature, Campante and Do (2008) and Parcero (2021) propose that in non-democratic countries, a large population and high population density lead to more redistributive policies. The reason is that larger population size and concentration potentially lead to a higher probability of unrest or revolution. As a preventative, governing elites may attempt to insure against political turmoil by implementing more redistributive policies.

Without ignoring the importance of national culture in achieving a knowledge economy and sustained growth (Khalil and Marouf, 2017), human capital has been a critical factor in the endogenous growth models. Most importantly, the accumulation of human capital may be easier for larger countries. On the one hand, Parcero (2021) suggests that larger non-democracies countries may provide a wider spread of primary and secondary education as a redistribution channel. Moreover, Alesina et al. (2020) suggest that governments use education as a way of indoctrination to homogenize the population, which is more compelling in larger countries. On the other hand, using university-level data, economies of scale in undergraduate teaching have been largely established (see, for instance, Cohn et al., 1989; Dundar & Lewis, 1995; Laband & Lentz, 2003).

Expenditure in R&D relative to GDP has been long considered a potential determinant of economic growth. See Lucas (1988) and Romer (1990) for pioneering endogenous incorporations of R&D into economic growth models and Das and Mukherjee (2020) for recent testing of this hypothesis. Moreover, larger countries may produce higher R&D relative to GDP, while smaller economies find it very difficult (Tiits et al., 2015). This is the case because R&D activities result in economies of scale (Hewitt, 1980; Hirschey and Caves, 1981) and economies of scope through the spillovers between different technology fields (Arora et al., 2011; Leten et al., 2007).

Many studies have attempted to assess empirically the effect of country size on economic performance (Robinson, 1960; Michaely, 1962; Pearson, 1965; Khalaf, 1974; Streeten, 1993; Damijan, 1996; Barro and Sala-i-Martin, 1995; Salvatore et al., 2001; Alesina et al., 2005). These studies typically find that small country size is associated with lower product diversification, higher trade openness, higher concentration of trade flows (in terms of both commodities and geographic destinations), larger governments, and larger balance of payments volatility. Yet the studies fail to find a significant relationship with country development levels (as measured by per capita GDP). Most recently, Rose (2006) takes a snapshot view of economic and socio-economic performance in large and small countries by considering various indicators. With the exception of trade openness, he could not confirm a significant size effect on any variables considered.

Lack of definitive empirical evidence in favor of the mainstream "large-is-better" view has generated alternative literature looking at the advantages of smallness. In an early study focusing on small countries, Easterly and Kraay (2000) found that, after controlling for location, small states have higher per capita GDP. They also reported that small countries do not have different per capita growth rates, though they are more volatile. The authors attribute the latter to small countries' higher exposure to trade shocks.

Hines (2005) finds that tax haven countries enjoy higher GDP per capita and higher GDP growth than those with higher taxes or more stringent tax enforcement. He claims that a distinctive feature of tax havens is their smallness: "The populations of seven of these countries exceeded 1 million in 1982, and these are referred to as the Big 7; other tax haven countries are known as Dots." (Hines 2005, p. 77). Blanco and Rogers (2011) also find evidence that tax haven policies positively affect economic growth. They go one step further and suggest that the observed favorable growth may be driven by factors related to size rather than to endogenous tax haven policies. This view finds support in the works of Bucovetsky and Haufler (2008), Kanbur and Keen (1993), and Winner (2005), whose models show that low tax rates may be advantageous for small countries.

As a consequence of lower tax rates, smaller countries can attract more FDI (Hines, 2005; Head and Ries, 2008). Indeed, Hines (2005) and Head and Ries (2008) find that FDI shares for small countries are more prominent than GDP shares. Finally, Anckar (2002) finds that micro-states or small island states do better in terms of democracy than the average country, though Srebrnik (2004) claims that evidence for this is not conclusive.

## 2.2.  Testable hypotheses

Our primary research question is whether there is a premium in country size. By this we mean whether there are significant gains from being a small or a large country in terms of certain socioeconomic indicators, and how large this premium is. However, based on the above review of theoretical and empirical literature, we make the following main hypotheses about the possible relationships between size and certain socioeconomic indicators:

1. market size enhances growth by raising specialization and intensity of product market competition, leading large countries to enjoy greater productivity and so faster long-term average rates of growth;
2. small countries are expected to have more difficulty handling adverse economic shocks because they cannot benefit from the help of other regions as it is the case in larger countries. As a result, they may show more unstable economic growth as measured by its standard deviation;
3. small countries are expected to be those that can afford to maintain the costs of smallness, implying that small countries will tend to be richer in terms of GDP per capita;
4. small countries are expected to have relatively larger governments; i.e., the share of government spending in GDP decreases as country size increases;
5. small countries are expected to have lower military spending relative to GDP and will be less likely to engage in armed conflict;
6. larger government spending relative to GDP, susceptibility to shocks and a lower ability to handle them may adversely affect smaller countries' public finance balances and levels of public debt;
7. small countries can compensate for their smallness by being more open to trade;
8. small countries are expected to receive more inflows of FDI, partly due to their lower taxation;

9. due to lower availability of resources as well as higher trade openness and net capital inflows, small countries are more likely to be subject to current accounts deficits;
10. large populations may lead to more redistributive policies and lower income inequality;
11. accumulation of human capital may be easier for larger countries, which implies that they are expected to show higher secondary and tertiary school enrollment and higher levels of public expenditure for education and R&D relative to GDP;
12. smaller countries tend to be more autocratic.

Using data on country performance for more than 200 countries in the period 1960 to 2015, in the following sections we will empirically test the validity of the above hypotheses and estimate the exact size premia in terms of various socio-economic indicators.

# 3. Empirical approach and data

This paper engages in an endeavor similar to that of Rose (2006), examining the impact of country size on socio-economic performance as proxied by a series of key indicators. Our approach, in addition to drawing on a larger dataset, diverges from the one used by Rose (2006) in two crucial ways. First, while Rose estimates the impact of size on country performance based on a continuous variable of size (population size), we use a semi-parametric approach, dividing countries into five size classes (micro, tiny, small, medium, and large), whose precise definitions are provided in the following subsection.

The reason for adopting this approach is that national economic structures and performance do not necessarily correspond to a continuous distribution of size as measured by population. Some public goods or types of production are indivisible by their nature, and require a certain threshold in terms of size. Excellent examples of this are independent national defense and national airspace monitoring, both prohibitively expensive for micro or tiny states. Increasing a country's population by a certain percent might not necessarily result in a linear increase in overall government expenditures on education or health, to mention only two areas. The provision of certain public goods may require a jump from a particular size class to another. Using a fully parametric approach makes it easy to overlook patterns in the data of the sort that we identify.

Ultimately, we are interested in showing whether there is a size premium in operation in various country performance indicators. And if so, what is the exact premium for being a micro, tiny, small, or either class of larger country, and for which indicators are the premia noteworthy?

## 3.1. Empirical approach

Our size premia are calculated from a regression of the log of socio-economic performance indicators on the corresponding categorical variable indicating size class and a set of control variables:

$$Y_{it} = \alpha + \beta \, size_i + \gamma \, control_{it} + \mu_t + \epsilon_{it}, \tag{1}$$

where $Y_{it}$ is a particular performance indicator for country $i$ in year $t$, in many cases expressed in logs. *Size* is defined as a time-invariant dummy variable taking value 1 if a country belongs to a certain size class, and zero otherwise. The variable $\mu_t$ controls for time-fixed effects. $\epsilon_{it}$ is the usual i.i.d error term.

Given that our primary variable of interest, size, is time-invariant, we cannot apply the standard fixed-effects estimation method. Thus, we estimate model (1) by OLS and we address the issue of potential correlation between the regressors and the error term due to unobserved country-fixed effects, to the extent possible, by adding a large number of time-invariant country-specific controls, which by nature are exogenous. Note that we provide also some robustness check by including size either as continuous variable (population) or various size dummies with cut-offs at different population size. However, as we are interested in estimating the size premia, we stick to the size variable measures based on size dummies, whereby we include time fixed effects and try to capture the time invariable country-fixed effects by including a rich set of country-specific control variables (see discussion below).

We use two different sets of size classes. In the first approach, *size* assumes a value of 1 if a country has a population of less than 15 million, and zero otherwise. The literature often suggests this threshold as the most appropriate. In the second approach, to tease out heterogeneity in the performance of our two broad country categories in terms of size, we refine the measure of size by allowing for five distinct size classes. A *micro* country has a population of less than 1 million. *Tiny*, *small*, and *medium-sized* countries have populations of between 1 and 5, 5 and 15, and 15 and 40 million, respectively. A country with more than 40 million is classified as *large*.

Two important notes regarding the construction of these country-size group dummies are in order. First, the dummy variables for each of the three smaller groups take the value of 1 if a country belongs to a given small-country group (micro, tiny or small) under consideration, a missing value if it belongs to one of the two other groups of small countries, and zero otherwise. This means that the comparison group for any of the three small-country groups is always the combined group of *medium* and *large* countries. We proceed similarly with the size dummy variables for groups of *medium* and *large* countries, where the comparison group is always the three small-country groups combined. Second, the size dummy variables are constructed by using a country's median value of population over the period 1960 to 2015. Thus, we prevent countries from potentially switching (by a small margin) among different size classes as their populations grow over time.

[*Insert Table 1*]

Table 1 shows that in 2015 three-quarters of countries (153 out of 200) belong to the broad category comprising the three small groups (with populations of less than 15 million). The largest number of countries in this category (53) are classified into the subgroup *micro*, followed by those in the subgroups *tiny* (52) and those in the subgroup *small* (48). The larger category is nearly equally comprised of *medium* (23) and *large* countries (24).[5]

---

[5] Note that we report only countries for which at least two variables (on population and GDP per capita) are available for a given year.

There are two sets of control variables used by estimating the model (1). Controls 1 include only time fixed effects. In addition to time-fixed effects, Controls 2 take into account a number of country-specific characteristics which serve as country-fixed effects and are exogenous to the size indicator. The most salient is the level of development (measured by the logarithm of real GDP per capita at the start of the period). The literature shows this to be the most important single determinant of a country's long-run performance. The second group of country characteristics consists of dummy variables indicating a country's geographic location. These variables include the log of the distance of a given country from the equator (in kilometers), binary dummy variables for landlocked and island nations, and regional dummies for developing countries (in Latin America, Sub-Saharan Africa, Middle East-North Africa, East Asia, South Asia, and Europe-Central Asia). The third group of characteristics controls for other income and cultural and historical characteristics of countries. Here, we include a dummy variable for High-Income countries (OECD), and language dummy variables for countries where English, French, German, Dutch, Portuguese, Spanish, Arabic, or Chinese is spoken. These language dummies control for the existence of similar cultural heritage. To control for a country's age and historical dependence ties, we include a binary dummy variable for countries created between 1800 and 1945, a dummy for countries created post-WW2, and a colonial dependency dummy—the latter controlling for long-lasting institutional effects that may result from prior colonial status. Finally, we also include a dummy variable for countries rich in natural resources. Here, we take into account the ongoing discussion in the literature whether natural resources are more of an economic curse than a blessing.[6] We follow the IMF classification, whereby IMF classifies 51 countries as "resource-rich." These are countries which derive at least 20% of exports or 20% of fiscal revenue from nonrenewable natural resources. Out of them, 22 countries are upper-middle- or high income, while 29 of these countries are low- and lower-middle-income.

There are four important notes to be made regarding the estimation of the model (1) and calculation of the size premia. First, our estimations of the model (1) include the complete set of control variables as explained above, including the natural resources variable. We provide some robustness tests below to show how including various time- and country-fixed effects affects the estimated coefficients.

Second, it is important to account for possible outliers in our data that may "*pollute the estimations*", which is justified as a general practice in the econometric literature. There are three mainly used approaches to do so. One is to exclude all observations in top and bottom 1 (or more) percentiles of the distribution of a particular variable. Second one is to use the Cook's distance method that measures the aggregate change in the estimated coefficients when each observation is left out of the estimation. Values of Cook's distance that are greater than 4/N may be problematic. And the third approach is to use robust regression specification in non- or semi-parametric empirical models as in these tests outliers won't necessarily violate their assumptions or distort their results.

---

[6] For instance, in two influential studies (Sachs and Warner, 1995, 2001) found a strong correlation between natural resource abundance and poor economic growth. However, recent studies find little support for the thesis. In a meta-study, Havranek et al (2016) find weak support for the thesis that resource richness adversely affects long-term economic growth. They note that "*approximately 40% of empirical papers finding a negative effect, 40% finding no effect, and 20% finding a positive effect*", but "*overall support for the resource curse hypothesis is weak when potential publication bias and method heterogeneity are taken into account.*" Kurtz and Brooks (2011) find that "*natural resource wealth can be either a "curse" or a "blessing" and that the distinction is conditioned by domestic and international factors, both amenable to change through public policy, namely, human capital formation and economic openness.*"

We addressed the issue of outliers in two ways. First, we identify outliers by using the Cook's distance method and exclude them from the regression analysis. Second, we apply robust regression analysis in our semiparametric estimations (size indicator and a number of the explanatory variables in the model have a nonparamateric form), which is one of the recommended methods to deal with problem of outliers.[7]

Third, although our dataset spans over the period 1960-2015, we actually estimate our model (1) only at 5-year intervals (i.e. for data in 1960, 1965 and so on until 2015). To avoid the noise and volatility in the data, for most of the variables we compute 5-year averages, such as average growth GDP, average current account deficit, average budget deficit, etc. This in turn means that we effectively deal with a panel data structure with a maximum of 12 observations per country at 5-year intervals. In this way we avoid the problem of possible cointegration between variables in our dataset structured as the panel data.

Finally, our estimations of model (1) are a basis for calculation of the size premia, i.e. of the premium for being a micro, tiny, small, or either class of larger country. A size premium shows the average percentage difference in performance between a particular country's size group and the reference country group. The size premium is calculated as follows. We first estimate model (1) separately for each dependent variable of interest and obtain a coefficient ß for a particular size indicator (whereby each coefficient corresponds to a separate regression). For cases in which the dependent variable takes a log form, we calculate the size premium from the estimated coefficient as $100*(\exp(\beta)-1)$ (where ß is the estimated size coefficient for a particular dependent variable). Otherwise, when the dependent variable cannot be expressed in logs, we estimate the size premia as $100(\beta/(\bar{y}/size))$, where *size* takes the value 1 and $\bar{y}$ is the mean of *y*.

### 3.2. Data

The dataset employed in this study consists of relevant country data sampled at five-year intervals, starting in 1960 and proceeding through 2015. Our dataset includes all 214 countries (or territories) taken from the World Bank's *World Development Indicators* (WDI). We follow Rose (2006) and consider a political territory to be a country if referred to as such by the WDI in 2015. In this way, our list includes several political territories that, strictly speaking, are not considered to be countries, such as the Cayman Islands, Hong Kong, Andorra, Puerto Rico, etc. Yet these entities have economies run by their autonomous authorities. Our list does not include countries that ceased to exist following the break-up during the period examined, such as the

---

[7] Excluding the outliers as described does have a minor impact on our results. For instance, in terms of the GDP per capita, excluding the identified outliers (18 out of 1,827 observations, i.e. about 1 per cent of all observations) does have a minor effect on improvement of the fit of regression (an improvement by 1.5 percentage points) as well as on the size of the main regressor (the latter changes at a third decimal point).

U.S.S.R., Czechoslovakia, or Yugoslavia. Instead, it includes their successor countries.[8]

Our primary source of data is WDI in 2015. For some indicators that were not available in the WDI, our source is the IMF's International Financial Statistics, Government Finance Statistics, Balance of Payments. The latter data mainly covers the period 1980 – 2015. Data on country characteristics, such as language, geographic location, and distance to the equator, was taken from Rose (2006). Data on conflict comes from the Uppsala Conflict Data Program (UCDP). Though WDI data collection began in 1960, data for some countries and indicators has become available only in more recent years. Table 2 shows summary statistics for data used. Though, our dataset includes all 214 countries, data for some countries are missing, which leads to an effective dataset of a maximum 200 countries in 2015 for some variables and less for other variables.

[*Insert Table 2*]

# 4. Results

In this section, we first present explorative graphic analysis for a bird's-eye view of the correlations between country size and selected variables. We then estimate size premia following the method already described.

## *4.1.  Bird's-eye view of the size effect*

Figure 1 shows scatter plots between the log of real GDP per capita and the log of population at five-year intervals from 1960 through 2010 and for the most recent year, 2012. Each point inside a graph represents a country. The line results from a simple bivariate linear regression, where the estimated coefficients represent simple elasticity of per capita GDP with respect to population size. Each picture also includes vertical (dashed) demarcation lines that separate the five size groups, running from left to right in this order: micro, tiny, small, medium, and large.

[*Insert Figure 1*]

Though the country sample increased over time, from 96 to 196, the relationship between per capita GDP and population size remains unaltered. The relationship between per capita GDP and size is weak but significantly negative, indicating that smaller countries are on average richer (more developed) than larger ones. Notice that the difference in per capita GDP can be quite high between the two extremes, i.e., between micro and large groups. This indicates that in terms of per capita income, there is a potentially significant negative size premium. We discuss exact size premia in the following subsection.

[*Insert Figure 2a*]
[*Insert Figure 2b*]

---

[8] See a list of countries in our dataset in Table A1 in Appendix.

Figures 2a and 2b explore the relationship between size and five-year averages of selected performance indicators for 1960 to 2015. The upper panel of Figure 2a shows that there is no relationship between average GDP growth and size. On the other side, unemployment seems to be (weakly) negatively related to population size, with a tendency of larger countries to exhibit lower unemployment rates. The lower panel of Figure 2a confirms that small countries insure against small domestic markets by engaging in international trade; there is evidence of a strong negative relationship between trade openness and size. Yet smaller countries do exhibit smaller current accounts surpluses.

Figure 2b graphically explores the relationship between population size and government size and the potential consequences of this relationship in terms of government budget balance and public debt. The upper panel of Figure 2b demonstrates that smaller countries, on average, have larger governments in terms of both expenditures and revenues. This confirms the notion that smaller nations may face higher per capita costs for certain public goods. Running larger governments, however, does not necessarily translate into fiscal irresponsibility. As shown in the lower panel of Figure 2b, there seems to be no significant relationship between fiscal stance or debt-to-GDP ratio and size.

The scatter plots representing relationships between population size and certain selected variables are instructive. However, these figures only show bivariate relationships and do not account for countries' heterogeneity along other dimensions.

One of the most critical factors influenced by population size is per capita GDP, whose correlation with the same set of selected performance variables is depicted in Figures 2c and 2d. The figures show that per capita income matters for country performance, and except for GDP growth, it matters a great deal. In particular, per capita income is negatively correlated with unemployment but positively correlated with trade openness and current accounts balances. Similarly, Figure 2d reveals that government size is strongly related to level of development. And surprisingly, so is government budget balance and public debt. Countries with higher per capita income seem to be more prudent in terms of fiscal policies. They run higher budget surpluses and lower levels of public debt.

[*Insert Figure 2c*]
[*Insert Figure 2d*]

We have made clear that size is correlated with per capita GDP and that both are correlated with performance indicators in a number of countries. This suggests that when searching for size effects, it is crucial to account also for income per capita effects and other country-specific covariates.

## 4.2. *Base empirical results*

This section presents the results from model (1), which serves as a basis for computing size premia. We only offer detailed results for the relationship between size and per capita GDP. For all the other variables, we simply show the computed size premia graphically. A visual presentation of the premia is far more instructive when considering many country performance indicators. The Appendix (Tables A3 and A4) offers tables with detailed results for all selected variables of interest.

Table 3 shows the results from estimating model (1) in a successive way for the variable GDP per capita. The first row corresponds to a bivariate regression of the log of GDP per capita (in 2005 constant dollars) on the Size dummy variable taking value 1 for population size smaller than 15 million, and 0 otherwise. The bivariate regression results in the first row show that countries with fewer than 15 million people experience significantly higher per capita income. The second row, labeled Controls 1, includes year fixed effects. The estimated coefficient becomes slightly smaller but remains significantly positive. The specification labeled Controls 2 includes the complete set of control variables as explained in Section 3.1 (with the obvious exception of GDP per capita). Under this specification, the size coefficient becomes more prominent and more significant. Subsequent rows running regressions year by year reveal that per capita income is always positively and significantly correlated with small size .

[*Insert Table 3*]

These results show that after controlling for a full set of country characteristics and time fixed effects, per capita GDP is significantly larger in countries with fewer than 15 million in population. Using the previously provided formula, the premium of smallness computed from the specification, with the full set of covariates, amounts to 58.2 percent. In other words, small countries, on average, experienced 58 percent higher income per capita than larger countries.

We also check for a robustness of the country size effects on GDP per capita using various size indicators. Similar to Rose (2006), we first employ population in the continuous form as a size indicator. As presented in Table 4, both OLS and fixed effects (FE) specifications show a negative correlation between GDP per capita and population size, whereby in the FE specification the size effect doubles as compared to the OLS specification. This confirms the importance of accounting for country-fixed effects. In the next three specifications, we employ three different country size indicators with a large country size cut-offs at 10, 15 and 20 million population. These specifications include complete set of country-specific fixed effects (Controls 2) as explained in Section 3.1. The results show that country size cut-offs yield similar results, with 15 million population cut-off yielding the highest fit and highest coefficient size.[9] The literature often suggests this threshold as the most appropriate.

[*Insert Table 4*]

Of course, these specifications with only one country size cut-off are very crude in accounting for country size. There might be differences across countries within these two country groups. We therefore run a set of regressions by classifying country size at more refined levels by defining five size classes – micro, tiny, small, medium, and large (see above for a definition of size classes). Regressions are run separately for each size class and include the complete set of control variables and time fixed effects. Table 5 shows the results.

[*Insert Table 5*]

The estimated coefficients by size groups in the pooled regression show that the three small country groups have significant size premia over larger countries in terms of GDP per capita. While the groups of tiny (population between 1 and 5 million) and small (population between 5 and 15 million) nations have significant premia of about 30 and 50 percent, respectively, over medium and large countries, real per capita income in 64 micro countries is, on average, more than 150 percent of that in medium and large countries. Looking at the other end of the scale, we observe that per capita income in the 24 medium and 25 large countries is, on average, lower than in the three groups of small countries combined, by 27 and 48 percent, respectively (see also Figure 3). Year-by-year regressions are affected by the smallness of the sample size, and hence only the coefficients for micro and large countries remain significant.

---

[9] See Table A2 in the Appendix for robustness checks also for all other variables in our analysis.

These results lead to one distinctive conclusion. After controlling for a large set of country-specific effects, including natural resources, small countries appear to have an easier time achieving higher standards of living, as measured by GDP per capita, than larger countries, despite their less diversified resources and smaller economic size. This is particularly true for the large group of micro countries.

### 4.3.    Size premia

This subsection presents results for size premia as computed from coefficients estimated using model (1). It starts with major economic indicators and follows with other socio-economic indicators.

### 4.3.1.  Size premia for economic indicators

Figures 3 to 3d show computed size premia for the following list of major economic indicators: GDP per capita, average GDP growth, standard deviation of GDP growth rate, average unemployment rate, average government revenues/GDP, average government expenditures/GDP, average current accounts/GDP, average openness, average budget balances/GDP, average debt/GDP, FDI/GDP, CPI, savings rate, investment rate, country risk, risk premia, and bank credit/GDP.

[*Insert Figure 3*]

Figure 3 presents size premia for the class of small countries as compared to large countries (the threshold is a population of 15 million). The figure demonstrates that small countries are significantly different from large ones: in particular, small countries perform significantly differently from large countries in 12 out of 17 indicators. This finding contrasts with that of Rose (2006), who uses parametric regression analysis to find that except for per capita income and trade openness, size does not matter.

The results show that small countries are on average richer by almost 60 percent, but they do not grow significantly faster. In other words, income per capita differences between small and large countries persist over time, which contrasts with implications stemming from a strict interpretation of endogenous growth theory. The lower intensity of product market competition in small countries does not necessarily lead to lower productivity benefits or lower long-term average growth rates.

On the other hand, economic growth in small countries is more volatile. The standard deviation of GDP growth rates over five-year intervals in small countries is on average 19 percent larger than in large countries. This is in line with the findings of Easterly and Kraay (2000), who also determine that growth is more volatile in smaller countries. In addition, it confirms the fourth potential scale effect put forward by Alesina and Spolaore (1997, 2003): small countries may have a lower ability to handle adverse economic shocks to which they are relatively more exposed due to their high trade openness when compared with large countries.

As shown by many studies, small countries are systematically and substantially more open than larger countries. Our results confirm these findings and additionally show that the small nation openness premium is close to 60 percent, i.e., shares of exports and imports in GDP in small countries are higher by almost 60 percent. In conjunction with the finding of no systematic differences in GDP growth, this is evidence in favor of the claim that small countries can compensate for the smallness of their domestic markets by becoming more open to trade. Another way in which small countries can become more open is in terms of extensive involvement in international capital flows. On average, small countries attract 40 percent more FDI relative to GDP, which is in line with the general findings of Hines (2005) and Head and Ries (2008). However, higher exposure to trade and FDI flows comes at the cost of systematically larger current accounts deficits (57 percent larger).

Small and large countries do not differ in terms of average unemployment rates but do differ in terms of inflation, where small countries on average exhibit 10 percent lower inflation. Small countries also exhibit lower saving and investment rates (both by five percent), whereby the savings rate premium is not significantly different from zero at ten percent. This premia difference between investment and saving rates indicates that small countries are dependent on foreign savings, which materialize in the form of larger net FDI and capital inflows and subsequently in higher current account deficits, as explained above.

As shown earlier, small countries tend to have bigger governments with average expenditure and revenue premia higher than those in larger countries (by 8 and 5 percent). While this confirms Alesina and Wacziarg's (1998) findings, having larger government does not necessarily lead to fiscal irresponsibility. On the contrary, small countries show more prudent fiscal policies and have budget surpluses larger by almost 160 percent in comparison with large countries. Moreover, there are no systematic differences between small and large countries in terms of public debt ratio to GDP. Larger budget surpluses and manageable public debt in small countries could be partly explained by the fact that small countries are more affluent, making it easier to collect taxes. Note that in all estimations, we control for GDP per capita, geographic location, the High Income (OECD) dummy, etc. Yet the significant effect of more prudent fiscal policies in small countries is still there.

Data also shows that the banking sector in small countries is close to 20 percent smaller on average than in large countries. Though small countries' risks are marginally lower from those of large countries, on average, small countries pay systematically higher risk premia (almost 40 percent) when taking out foreign loans. At least here, small countries are taxed for their smallness by international financial markets. The reason may well lie in perceptions that notwithstanding more sound fundamentals such as better budget positions, small countries are perceived as having a lesser ability to repay loans in the long run.

[*Insert Figure 3a*]
[*Insert Figure 3b*]
[*Insert Figure 3c*]
[*Insert Figure 3d*]

We now show that the differences described above between small and large countries are generally robust to more refined groups of small countries. In particular, in terms of per capita GDP, micro countries (with population below 1 million) clearly outperform the group of larger countries (i.e., medium and large countries combined) with a premium of close to 130 percent. Tiny and small countries show smaller but still significant premia of about 50 and 20 percent, respectively. A look at the other end of the scale shows that medium and large countries are poorer than the three groups of small countries, by almost 50 percent for large and 30 percent for medium-sized countries.

In terms of GDP growth, there are no significant differences in any of the three groups of smaller countries. However, significant differences appear in the club of smaller countries in terms of GDP growth volatility. While the group of small countries (with population between 5 and 15 million) shows no difference in growth volatility compared to larger countries, the groups of tiny and micro countries do experience more volatility. Growth volatility in tiny and micro countries is 20 and almost 50 percent above that of larger countries, respectively. Hence smaller countries are more vulnerable to shocks in economic activity.

In terms of trade openness, the differences among groups of smaller countries are quite pronounced. In the group of small countries, the premium is more than 40 percent, but goes up to almost 80 percent in the group of micro countries. The trends are similar in terms of FDI to GDP ratios. However, the notable difference is that the most prominent premium over larger countries appears in the group of tiny countries (i.e., 70 percent). FDI premia in small and micro countries amount to 20 and 15 percent, respectively. Current accounts deficits seem to be the greatest plague in tiny and small countries. Surprisingly, current accounts deficits are not significantly different between large countries and micro countries. There seems to be no convenient explanation for this phenomenon.

In terms of unemployment, no systematic similarities emerge across country groupings. Relatively to large countries, micro nations are subject to significantly lower unemployment rates, while for the groups of tiny and small countries, there is no systematic difference. In terms of inflation (CPI), the group of tiny countries is driving the overall result of lower inflation in the broad group of small countries. For all other country groups there are no significant differences from the corresponding control groups. Medium-sized countries experience higher unemployment rates (the premium over smaller countries is 17 percent).

Government size, budget balancing and size of banking sectors are evidently systematically related to country size, whereby the biggest premia apply to the both extremes of the size distribution – to micro and to large countries. The smaller the country, the bigger the government in relative terms, the bigger the budget surplus, and the smaller the banking sector. However, also micro and small countries enjoy more sound fiscal policies, while the group of small countries enjoys lower debt levels.

In terms of saving and investment rates there seem to be a clear pattern of association between the two and the country size. However, most of the premia at the more refined country groups are not significantly different from zero and there seem to be two country groups that are driving the overall results. Lower savings rates are driven by the group of small countries, while higher investment rates are driven by the group of medium-sized countries.

Finally, in terms of the risk premium on foreign borrowing, differences across country groups are very systematic. It is micro countries that pay the highest tax on smallness in terms of higher risk premium (about 130 percent), while medium and large countries pay the lowest premia (by 25 and 30 percent, respectively).

### 4.3.2. Size premia for socio-economic indicators

Finally, we turn to other socio-economic indicators of interest, such as education expenditures as percentages of GDP, secondary school enrollment, tertiary school enrollment, human development index, life expectancy at birth, public health expenditures as percentages of GDP, infant mortality, internet users, mobile phone subscribers, fixed telephone lines, road density, military expenditure to GDP, number of homicides, number of armed conflicts, democracy index, autocracy index, ease of doing business index, informal payments, Gini index and Kuznets ratio of inequality and additional indicators of knowledge economy, such as high technology share in manufacturing exports, share of R&D expenditures in GDP and patents per thousand of population.

*[Insert Figure 4a]*
*[Insert Figure 4b]*

Figures 4a and 4b present size premia for a class of small countries compared to large countries (the dividing line is 15 million). These figures demonstrate several interesting findings. First, on average, public spending for health care relative to GDP is higher in small countries by almost 10 percent. However, overall quality of life (as measured by the human development index) and life expectancy are not significantly higher (both insignificant from zero at ten percent), while infant mortality is higher (by eight percent) in small countries. As shown in Figure 5b, this is due mainly to the low performance of micro countries, and to a lesser extent, tiny countries. The group of small countries does not perform differently from the two groups of larger countries. Relatedly, Amate-Fortes et al. (2017) find that island countries do better in terms of human development index, but being an island is not equivalent to being small in terms of population.

Second, despite higher expenditures for public education (by five percent), secondary and tertiary school enrolment rates are either equal or significantly lower in small countries (with negative premia for the latter at almost eight percent). Again, this is predominantly due to low performance by micro-sized countries (see Figure 5c).

Third, smaller countries have less telecommunication infrastructure (as measured by internet users, mobile phone subscribers, and fixed telephone lines). Here again, micro and tiny countries perform more poorly, but the negative premium is also evident in the group of small countries. These findings confirm Alesina and Spolaore's (1997, 2003) proposition that the cost of providing public goods in smaller countries is higher, resulting in their lower provision and lower overall quality.

*[Insert Figure 5a]*
*[Insert Figure 5b]*
*[Insert Figure 5c]*
*[Insert Figure 5d]*
*[Insert Figure 5e]*

Fourth, in contrast to propositions in the literature, our findings indicate that small countries do not seem to be adversely affected to any significant degree in terms of military spending (see Figures 4b and 5d). Though, there are differences among the groups of small cuntries. In the smallest, micro countries, military expenditure relative to GDP is lower than in the two groups of large countries (by 15 percent), while in the group of small countries with population between 5 and 15 million military spendings are significantly higher than in large countries. There seem to be two tendencies at work. First, the combination of lower military spending and higher wealth is compatible with previous findings (Desli and Gkoulgkoutsika, 2020). On the one hand, this might be a consequence of the lesser tendency of smaller countries, particularly micro countries, to build sophisticated military systems. At least in this case, the higher per capita cost of providing public goods favors smaller countries. On the other hand, smaller countries tend to adopt less aggressive stances towards other nations, as shown by their significantly lower engagement in armed conflicts. On average, the group of small countries engages in armed conflicts almost by 20 percent less frequently than the two groups of larger countries.

Fifth, Figures 4b and 5c show small countries to be safer, which is reflected in a negative, though not significant, association between size and number of intentional homicides. This is consistent with Zanchi et al. (2021) finding in a study on 70 countries from 2010 to 2015.

Sixth, the results in Figure 4b show that small countries are more prone to autocratic tendencies as shown in two alternative measures of democracy: the Democracy index and the Autocracy index. Small countries tend to be less democratic (with a negative premium of 20 percent relative to the two groups of large countries) and more autocratic, though the latter premia is not significant at ten percent. A closer look reveals that these tendencies are driven by the group of countries with extremely low population size (micro countries) (see Figure 5d). This is in line with findings in the literature (see Barro, 1999; Rose, 2006). Most recently, Erisen and Wiltse (2017) find a negative association between country size and democracy. They measure democracy by using the variable Polity2 taken from the Polity IV dataset) for 163 states in the period 1960-2012.

Seventh, the finding about democracy and autocracy is somewhat related to another hypothesis in the literature (see Campante and Do, 2008; Parcero, 2021). In non-democracies, a larger population leads to more redistributive policies, likely to head off revolts against ruling elites. Moreover, independently of the democratic status, this relationship is also found for 43 countries from 1991 to 2016 (Lee and Wang, 2021). Our results show no systematic differences among large and small countries in terms of inequality. Note, however, that this is not a contradiction to the referenced hypothesis, because our sample also includes democracies. In particular, we find that two standard measures of income inequality (the Gini index and Kuznets ratio)[10] show no systematic correlation with country size (see Figures 4b and 5e). The only exception is the group of small countries (population range of 5 to 15 million). They display a significantly smaller Gini index (by four percent) relative to larger countries and a similarly smaller, though not significant, Kuznets ratio.

---

[10] The Kuznets ratio is defined as the ratio of income shares between the highest 20% of earners and the lowest 40% earners.

Eight, results demonstrate that small countries do not lag in providing sound business environments. The group of tiny countries enjoys a significantly sounder business environment as measured by the World Bank Doing Business ranking than larger groupings. This confirms the high rankings of smaller countries on IMD World Competitiveness Center (2012) and WEF competitiveness scoreboards.[11]

Finally, smallness seems to be a handicap in the knowledge economy. Small countries invest less in R&D relative to GDP, with the exception of the subset of tiny countries, and they systematically underperform in terms of output - patents filed per thousand of population and share of high-tech exports. For patents, the negative premium for small countries ranges from 2 to 4 percent, while for the share of high-technology exports it ranges from 25 to 75 percent, compared with the group of large countries. This novel results may be related to the fundamental nature of endogenous growth theory, according to which country size matters for knowledge spillovers and their translation into innovation, patents, and exports of knowledge-intensive products.

# 5. Conclusions

Using an econometric approach, we tease out premia related to small country population size in a variety of key dimensions. We find that after controlling for several country-specific characteristics, small countries behave in distinct ways, and differ from larger ones. The most salient findings indicate that small countries are richer and have larger governments but are also more prudent in fiscal policies and run smaller public debts. In terms of economic performance, small countries generally do better than large countries, compensating for smallness by relying on foreign trade and FDI. This is particularly true for micro-sized countries. Smallness does not result in higher income inequality or less democracy. It comes as no surprise then that country splintering and attempts of regions to secede from larger wholes have been so widespread during the twenty century, Coggins (2011). The current findings may offer a complementary explanation as to why this has been the case.

The generally better performance of smaller nations comes at the cost of higher vulnerability to external shocks, resulting in higher volatility of growth rates. Smaller countries seem to pay higher absolute and per capita costs to provide essential public goods, but get less for their penny in health and education outcomes. The same is not true for military spending, where small countries exhibit lower spending and a lesser tendency to engage in armed conflict.

This study is complementary to that of Rose (2006). It covers a period fifteen years longer and confirms his results in many ways, but reaches some different conclusions. In particular, our non-parametric approach indicates that the relationship between some economic and socio-economic indicators is not linear. This is true for unemployment rates, inflation, debt to GDP, investment rates, road density, life expectancy, infant mortality, secondary school enrollment, military expenditure to GDP, democracy index, ease of doing business, and income inequality, to mention a few. Our results also show that the break in linearity is mostly produced at the intersection between medium and large countries, particularly concerning socio-economic indicators. However, non-linearities at the intersection between micro and tiny countries and to a lesser extent between tiny and small countries are also observed.

---

[11] Note that due to the lower rate of coverage of countries both in IMD World Competitiveness Center (2012) and WEF scoreboards we are unable to use them in our empirical analysis.

This paper's findings offer essential guidance to policymakers, international organizations, and business researchers responsible for assessing countries' economic or socio-economic performance or potential. Our study implies that a comparison with larger nations may be of little help in assessing how to spur economic success is smaller nations. Instead, a more fruitful approach would be to look at a group of countries sharing similar characteristics, such as size. For micro countries, the size premium is considerable across many indicators.

# Tables to be included in the text

Table 1: Distribution of countries by population size

| | | | 1960 | 1965 | 1970 | 1975 | 1980 | 1985 | 1990 | 1995 | 2000 | 2005 | 2010 | 2015 |
|---|---|---|---|---|---|---|---|---|---|---|---|---|---|---|
| micro | pop <1 | no. | 11 | 13 | 23 | 26 | 36 | 42 | 46 | 48 | 50 | 53 | 53 | 53 |
| | | pop. (mill.) | 0.147 | 0.162 | 0.179 | 0.196 | 0.213 | 0.237 | 0.266 | 0.291 | 0.310 | 0.343 | 0.391 | 0.409 |
| tiny | 1 < pop < 5 | no. | 23 | 25 | 25 | 27 | 32 | 35 | 42 | 48 | 51 | 52 | 52 | 52 |
| | | pop. (mill.) | 1.7 | 2.0 | 2.2 | 2.4 | 2.6 | 2.9 | 3.2 | 3.4 | 3.7 | 3.9 | 4.3 | 4.4 |
| small | 5 < pop < 15 | no. | 28 | 30 | 33 | 33 | 36 | 38 | 44 | 46 | 46 | 47 | 47 | 48 |
| | | pop. (mill.) | 5.3 | 5.7 | 6.3 | 6.9 | 7.8 | 8.5 | 9.3 | 10.1 | 11.1 | 12.3 | 13.5 | 13.2 |
| medium | 15 < pop < 40 | no. | 16 | 16 | 16 | 16 | 17 | 18 | 22 | 22 | 23 | 23 | 23 | 23 |
| | | pop. (mill.) | 13.3 | 14.9 | 16.6 | 18.6 | 20.8 | 22.5 | 25.0 | 27.4 | 29.4 | 31.8 | 34.3 | 35.4 |
| large | 40 < pop | no. | 18 | 19 | 20 | 20 | 20 | 22 | 24 | 24 | 24 | 24 | 24 | 24 |
| | | pop. (mill.) | 111.9 | 117.4 | 128.6 | 142.3 | 155.2 | 158.3 | 166.7 | 179.1 | 190.9 | 201.6 | 217.8 | 222.1 |

Note: The table includes countries in the analysis for which at least two variables (on population and GDP per capita) are available for a given year.

Source: *World Development Indicators*, World Bank.

Table 2: Variables and descriptive statistics of the sample data (2015)

| Variable | Obs | Mean | Std. Dev. | Min | Max |
|---|---|---|---|---|---|
| Average annual GDP growth over past 5 years | 1,646 | 1.8 | 3.7 | -22.3 | 31.5 |
| Autocracy Index | 625 | 3.8 | 3.7 | 0.0 | 10.0 |
| Domestic credit provided by banking sector (% of GDP) | 1,591 | 72.6 | 508.8 | -67.5 | 15,676.0 |
| Current account balance (% of GDP) | 1,340 | -2.9 | 12.2 | -90.0 | 106.8 |
| Passenger cars (per 1,000 people) | 200 | 234.3 | 214.3 | 1.0 | 1,139.1 |
| Mobile cellular subscriptions (per 100 people) | 2,376 | 21.6 | 42.1 | 0.0 | 284.3 |
| No. of armed conflict | 608 | 0.3 | 0.5 | 0.0 | 1.0 |
| Inflation, consumer prices (annual %) | 1,457 | 33.3 | 386.3 | -17.6 | 11,749.6 |
| Central government debt, total (% of GDP) | 862 | 60.2 | 55.3 | 0.0 | 755.3 |
| Budget surplus/deficit (% of GDP) | 1,001 | -1.9 | 7.1 | -34.2 | 51.4 |
| Democracy Index | 625 | 3.9 | 4.2 | 0.0 | 10.0 |
| Log Km from equator | 2,460 | 2857.0 | 1904.0 | 0.0 | 8,015.4 |
| Ease of doing business index | 184 | 93.4 | 53.4 | 1.0 | 185.0 |
| Public spending on education, total (% of GDP) | 689 | 4.4 | 2.0 | 0.3 | 14.8 |
| Foreign direct investment, net inflows (% of GDP) | 4,650 | 4.0 | 8.5 | -161.0 | 173.0 |
| Gross fixed capital formation (% of GDP) | 1,502 | 21.8 | 8.3 | 1.1 | 92.4 |
| Firms expected to give gifts to tax officials (% of firms) | 74 | 25.0 | 24.3 | 0.0 | 83.8 |
| Gini Index | 179 | 41.6 | 9.9 | 24.2 | 62.8 |
| Central gov budget expenditures (% of GDP) | 954 | 32.4 | 13.4 | 0.6 | 104.1 |
| Central gov budget revenues (% of GDP) | 960 | 30.4 | 13.6 | 0.0 | 98.5 |
| Human Development Index | 367 | 0.7 | 0.2 | 0.3 | 1.0 |
| Health expenditure, public (% of GDP) | 749 | 3.7 | 2.4 | 0.0 | 19.3 |
| Intentional homicides (per 100,000 people) | 376 | 9.8 | 14.7 | 0.0 | 139.1 |
| Hospital beds (per 1,000 people) | 1,016 | 4.5 | 3.9 | 0.1 | 40.3 |
| High technology exports share (% manufacturing exports) | 278 | 10.1 | 11.1 | 0.0 | 52.3 |
| Country risk | 138 | 68.1 | 11.6 | 34.8 | 90.8 |
| IMD Comp Index | 46 | 24.0 | 13.9 | 1.0 | 47.0 |
| Income share held by highest 10% | 785 | 33.3 | 7.8 | 18.2 | 65.0 |
| Income share held by highest 20% | 785 | 48.7 | 8.2 | 31.4 | 78.3 |
| Income share held by lowest 20% | 785 | 5.8 | 2.3 | 0.0 | 11.9 |
| Income share held by second 20% | 784 | 10.0 | 2.5 | 1.9 | 15.8 |
| Mortality rate, infant (per 1,000 live births) | 2,021 | 54.4 | 47.2 | 1.7 | 249.4 |
| Informal payments to public officials (% of firms) | 76 | 23.7 | 18.1 | 0.0 | 69.9 |
| Internet users (per 100 people) | 1,176 | 18.1 | 25.6 | 0.0 | 96.0 |
| Kuznets ratio (highest 20 to lowest 40 % of income) | 784 | 366.5 | 233.3 | 113.4 | 3747.3 |
| Life expectancy at birth, total (years) | 2,099 | 62.6 | 11.8 | 30.3 | 83.2 |
| Military expenditure (% of GDP) | 681 | 2.6 | 3.1 | 0.1 | 48.7 |
| Trade (% of GDP) | 1,678 | 79.5 | 50.7 | 1.1 | 447.2 |
| Patents applied by residents (per million people) | 688 | 0.1 | 0.3 | 0.0 | 3.0 |
| Telephone lines (per 100 people) | 2,089 | 14.2 | 18.0 | 0.0 | 125.5 |
| Political Stability, KKZ | 164 | 0.0 | 1.0 | -2.8 | 1.7 |
| Population (million) | 2,540 | 23.9 | 99.5 | 0.0 | 1,350.0 |
| Poverty gap at national poverty line (%) | 206 | 12.2 | 10.4 | 0.4 | 47.6 |
| R&D expenditure share (% GDP) | 337 | 0.9 | 0.9 | 0.0 | 4.2 |
| GDP per capita (constant 2005 US$) | 1,847 | 8,938 | 14,771 | 50 | 127,000 |
| GDP per capita, PPP (constant 2005 international $) | 1,315 | 10,637 | 12,886 | 102 | 123,000 |
| Risk premium on lending ( %) | 570 | 6.5 | 15.1 | -4.3 | 293.3 |
| Road density (km of road per 100 sq. km of land area) | 180 | 111.5 | 319.5 | 1.0 | 3,850.0 |
| Rule of Law | 184 | 0.0 | 1.0 | -2.3 | 2.2 |
| Gross domestic savings (% of GDP) | 1,599 | 17.5 | 16.4 | -86.9 | 85.6 |
| Standard deviation of GDP growth over past 5 years | 1,834 | 3.7 | 3.6 | 0.0 | 48.8 |
| School enrollment, secondary (% gross) | 1,116 | 62.1 | 33.8 | 0.6 | 161.7 |
| Tax revenue (% of GDP) | 444 | 16.6 | 7.8 | 0.2 | 60.8 |
| School enrollment, tertiary (% gross) | 974 | 21.4 | 21.5 | 0.0 | 103.1 |
| Unemployment, total (% of total labor force) | 823 | 8.7 | 6.0 | 0.0 | 41.4 |
| Voice and Accountability, KKZ | 188 | 0.0 | 1.0 | -2.1 | 1.6 |
| WEF Competitiveness Index | 56 | 29.2 | 16.8 | 1.0 | 58.0 |

Source: *World Development Indicators*, World Bank; *International Financial Statistics, Government Finance Statistics, Balance of Payments*, IMF; Rose (2006), *Uppsala Conflict Data Program* (UCDP).

Table 3: Per capita GDP and size (small threshold is at 15 million)

|  | Size | t-stat | Obs. | R-sq. |
|---|---|---|---|---|
| Bivariate | 0.228 | [2.72]*** | 1,847 | 0.004 |
| Control 1 | 0.194 | [2.33]** | 1,847 | 0.023 |
| Control 2 | 0.459 | [9.95]*** | 1,827 | 0.805 |
| 1960 | 0.309 | [1.85]* | 96 | 0.884 |
| 1965 | 0.296 | [1.70]* | 103 | 0.871 |
| 1970 | 0.406 | [2.47]** | 117 | 0.860 |
| 1975 | 0.514 | [2.77]*** | 122 | 0.852 |
| 1980 | 0.417 | [2.56]** | 141 | 0.861 |
| 1985 | 0.553 | [3.55]*** | 155 | 0.851 |
| 1990 | 0.476 | [3.09]*** | 176 | 0.833 |
| 1995 | 0.380 | [2.26]** | 186 | 0.796 |
| 2000 | 0.459 | [2.99]*** | 190 | 0.810 |
| 2005 | 0.448 | [3.01]*** | 195 | 0.800 |
| 2010 | 0.436 | [2.91]*** | 195 | 0.783 |
| 2015 | 0.346 | [2.20]** | 196 | 0.773 |

*Notes*: Results of estimating model (1). Coefficients of regressions of log GDP per capita (2005 constant $) on Size dummy variable taking the value of 1 for population size smaller than 15 million, and 0 otherwise. Each row represents a separate regression. Set of control variables in Controls 1 includes year fixed effects only. Controls 2 consists of the complete set of control variables as explained in Section 3.1. Control variables and the constant term are omitted from the presentation for brevity.

Robust t-statistics in brackets; *** $p<0.01$, ** $p<0.05$, * $p<0.1$.

Table 4: Robustness check for size indicators (GDP per capita)

| Size indicator | (1) log (Pop) OLS | (2) log (Pop) FE | (3) Size 10 OLS | (4) Size 15 OLS | (5) Size 20 OLS |
|---|---|---|---|---|---|
| Log Population | -0.257 [-21.73]*** | -0.546 [-11.16]*** |  |  |  |
| Size 10 mill. |  |  | 0.376 [8.10]*** |  |  |
| Size 15 mill. |  |  |  | 0.459 [9.95]*** |  |
| Size 20 mill. |  |  |  |  | 0.432 [9.50]*** |
| Observations | 1,827 | 1,827 | 1,827 | 1,827 | 1,827 |
| R-squared | 0.842 | 0.971 | 0.803 | 0.805 | 0.803 |

*Notes*: Results of estimating model (1). Coefficients of regressions of log GDP per capita (2005 constant $) on indicated size indicator. Each row represents a separate regression. Model is estimated for the pooled sample and includes complete set of control variables (Controls 2) as explained in Section 3.1. Control variables and the constant term are omitted from the presentation for brevity.

Robust t-statistics in brackets; *** $p<0.01$, ** $p<0.05$, * $p<0.1$.

Table 5: Per capita GDP and size (by size classes)

| | micro | | tiny | | small | | medium | | large | | Obs. |
|---|---|---|---|---|---|---|---|---|---|---|---|
| Pooled | 0.942 | [10.21]*** | 0.401 | [7.61]** | 0.271 | [4.84]** | -0.317 | [-5.43]** | -0.659 | [-11.73]** | 1,827 |
| 1960 | 1.019 | [2.51]** | 0.016 | [0.08] | 0.179 | [0.91] | -0.064 | [-0.33] | -0.653 | [-2.85]*** | 96 |
| 1965 | 1.136 | [2.82]*** | 0.068 | [0.38] | 0.203 | [1.08] | -0.120 | [-0.53] | -0.544 | [-2.37]** | 103 |
| 1970 | 0.944 | [2.66]** | 0.222 | [0.98] | 0.288 | [1.58] | -0.178 | [-0.89] | -0.651 | [-2.92]*** | 117 |
| 1975 | 1.228 | [3.02]*** | 0.251 | [1.30] | 0.310 | [1.41] | -0.374 | [-1.56] | -0.688 | [-2.88]*** | 122 |
| 1980 | 0.807 | [2.21]** | 0.261 | [1.37] | 0.265 | [1.25] | -0.260 | [-1.29] | -0.619 | [-3.01]*** | 141 |
| 1985 | 1.113 | [3.54]*** | 0.509 | [2.29]** | 0.385 | [1.96]* | -0.358 | [-1.90]* | -0.753 | [-3.77]*** | 155 |
| 1990 | 0.949 | [3.12]*** | 0.457 | [2.38]** | 0.397 | [2.04]** | -0.313 | [-1.65] | -0.658 | [-3.49]*** | 176 |
| 1995 | 1.017 | [3.15]*** | 0.343 | [1.66] | 0.234 | [1.13] | -0.226 | [-1.07] | -0.561 | [-2.68]*** | 186 |
| 2000 | 1.003 | [3.31]*** | 0.449 | [2.49]** | 0.250 | [1.21] | -0.324 | [-1.62] | -0.643 | [-3.36]*** | 190 |
| 2005 | 1.041 | [3.32]*** | 0.419 | [2.28]** | 0.218 | [1.11] | -0.326 | [-1.63] | -0.607 | [-3.42]*** | 195 |
| 2010 | 0.879 | [2.69]*** | 0.347 | [1.72]* | 0.212 | [1.11] | -0.372 | [-1.78]* | -0.561 | [-3.23]*** | 195 |
| 2015 | 0.768 | [2.27]** | 0.288 | [1.41] | 0.103 | [0.55] | -0.280 | [-1.27] | -0.458 | [-2.47]** | 196 |

*Notes*: Results of estimating model (1). Coefficients of regressions of log GDP per capita (in 2005 constant dollars) on five Size dummy variables. Each coefficient corresponds to a separate regression. Regressions include the complete set of control variables as explained in Section 3.1. Control variables and the constant term are omitted from the presentation for brevity.

Robust t-statistics in brackets; *** p<0.01, ** p<0.05, * p<0.1.

**Figures to be included in text**

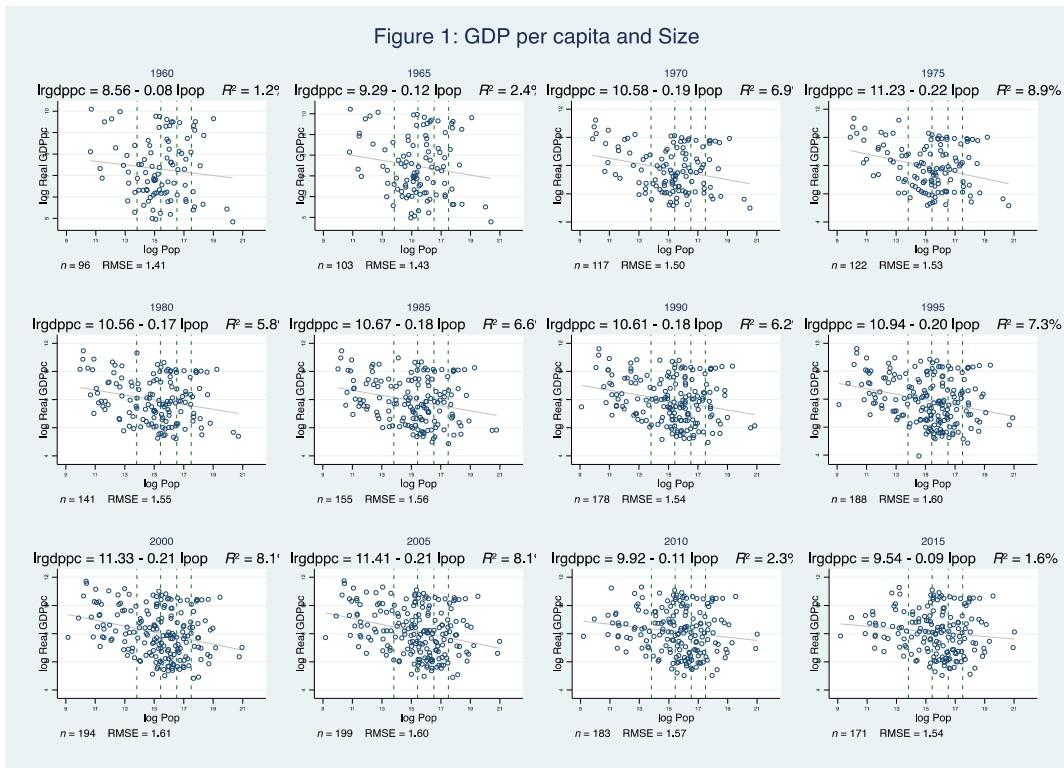

Figure 1 : GDP per capita and Size

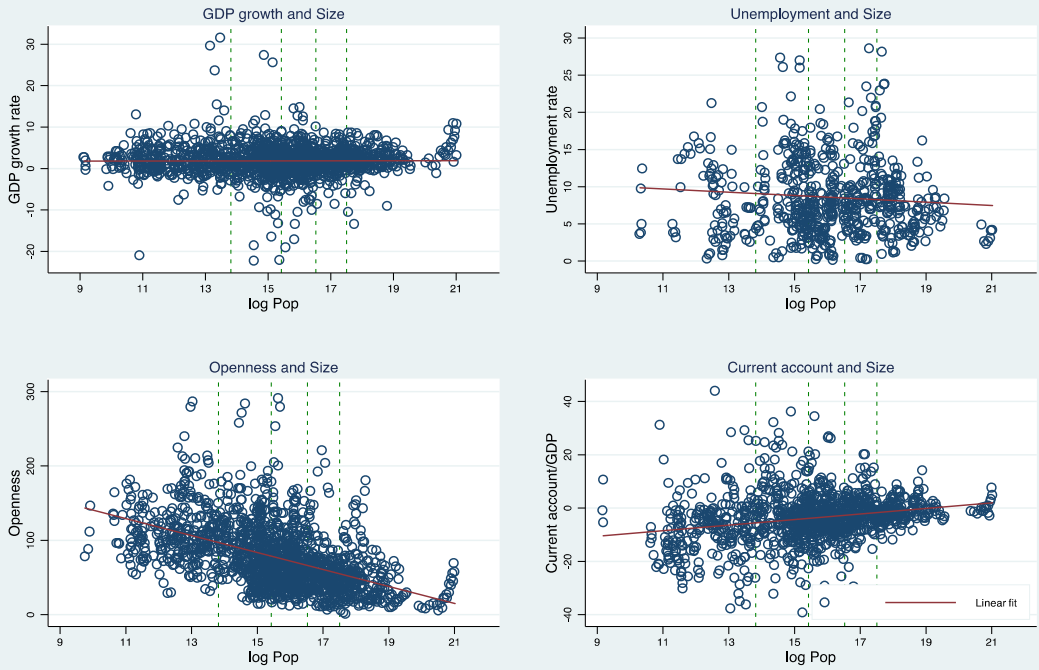

Figure 2a: Does size matter?

Note: Dashed lines indicate upper boundaries of country size classes

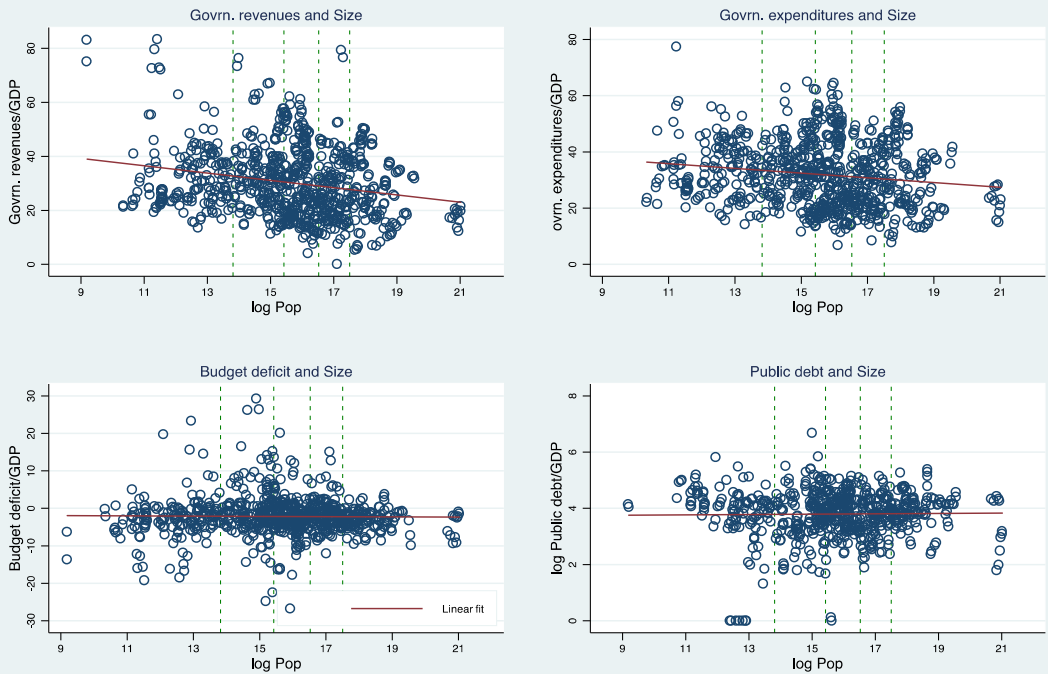

Figure 2b: Does size matter?

Note: Dashed lines indicate upper boundaries of country size classes

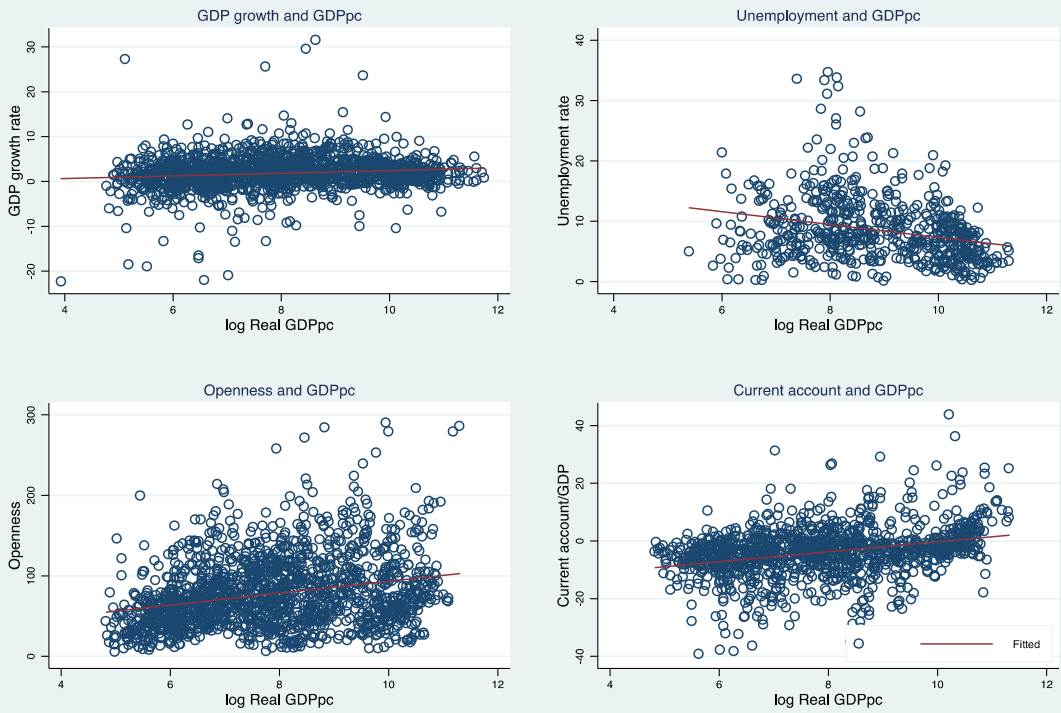

Figure 2c: Does development matter?

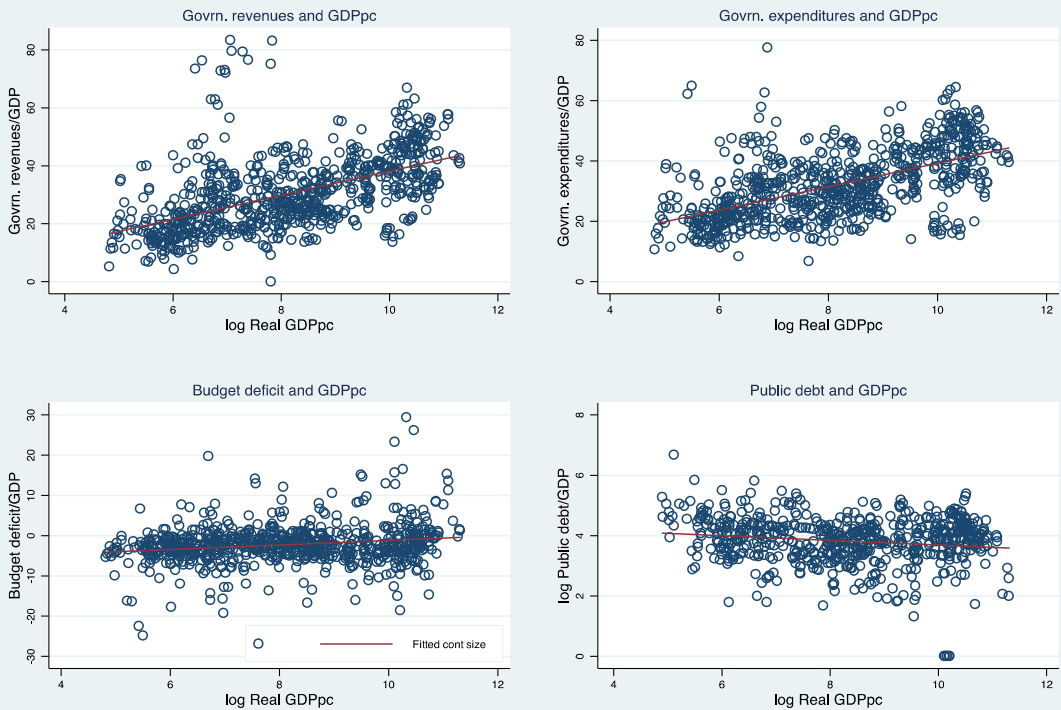

Figure 2d: Does development matter?

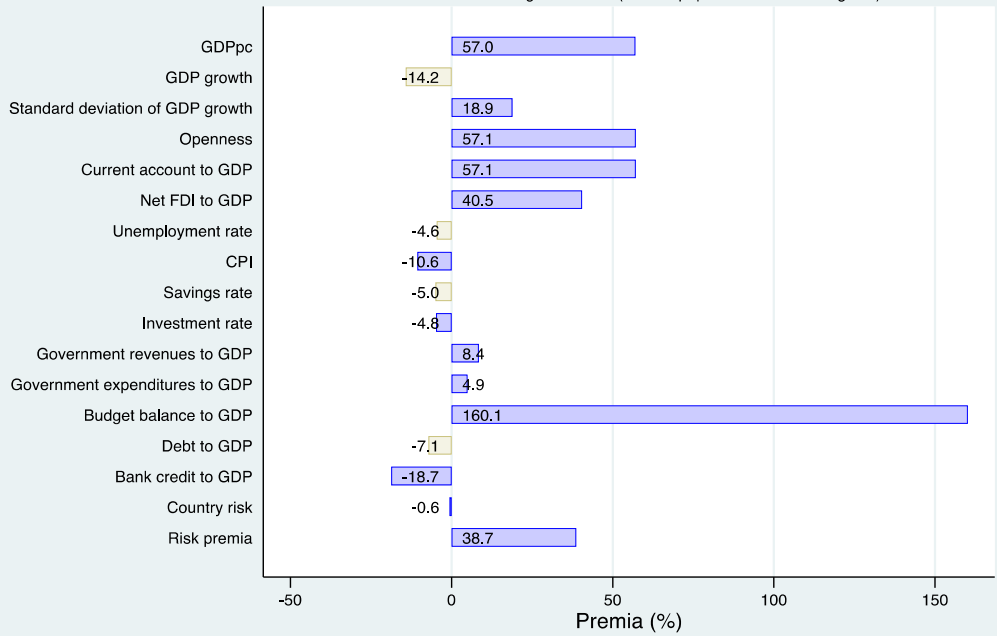

Figure 3: Size premia (%)

Premia of small v. large countries (15 mill. population as a dividing line)

Note: blue bar: significant at 10%; yellow bar: insignificant at 10%

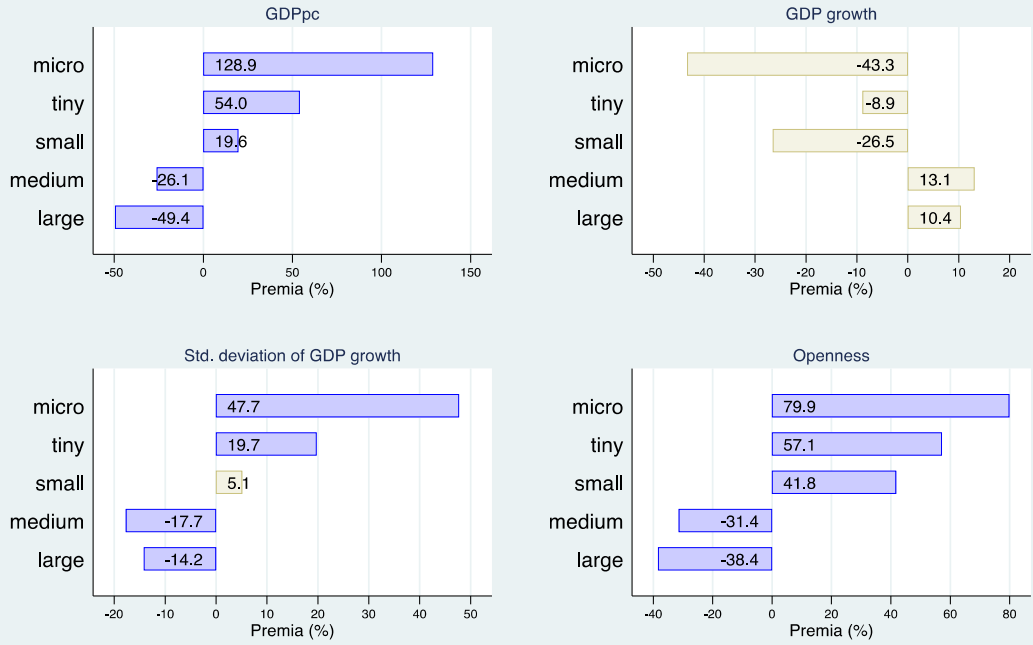

Figure 3a: Size premia (%)

Premia over group of small or large countries (by five country size groups)

Note: blue bar: significant at 10%; yellow bar: insignificant at 10%

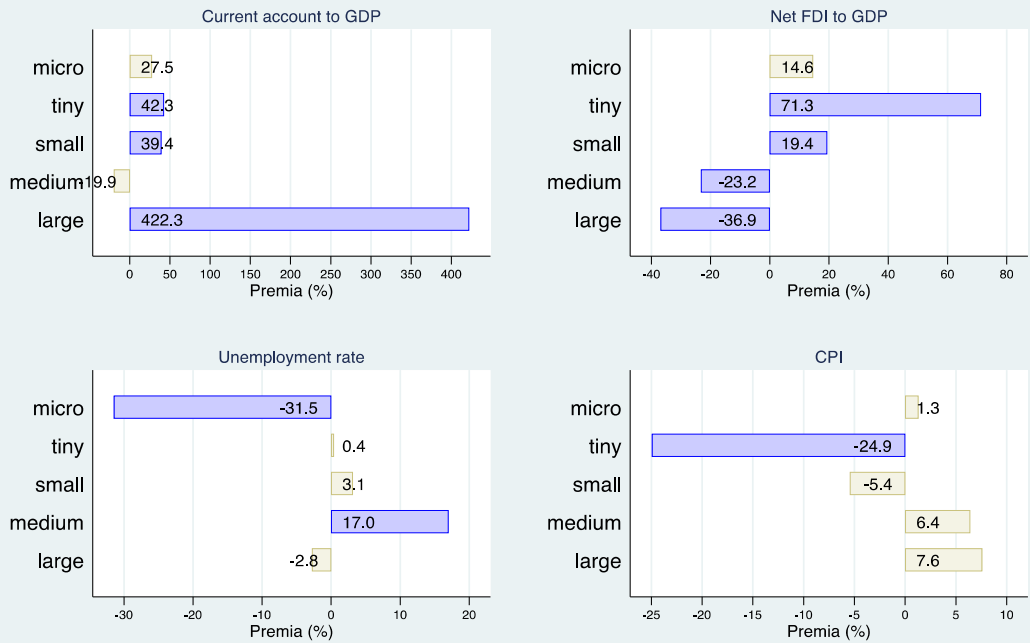

Figure 3b: Size premia (%)

Premia over group of small or large countries (by five country size groups)

Note: blue bar: significant at 10%; yellow bar: insignificant at 10%

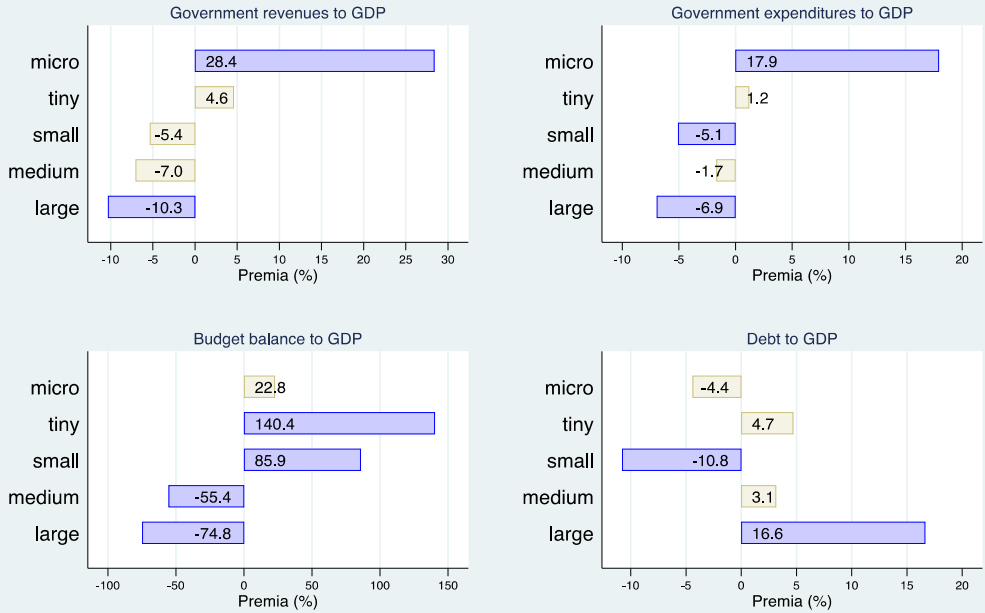

**Figure 3c: Size premia (%)**

Premia over group of small or large countries (by five country size groups)

Government revenues to GDP

| | |
|---|---|
| micro | 28.4 |
| tiny | 4.6 |
| small | -5.4 |
| medium | -7.0 |
| large | -10.3 |

Government expenditures to GDP

| | |
|---|---|
| micro | 17.9 |
| tiny | 1.2 |
| small | -5.1 |
| medium | -1.7 |
| large | -6.9 |

Budget balance to GDP

| | |
|---|---|
| micro | 22.8 |
| tiny | 140.4 |
| small | 85.9 |
| medium | -55.4 |
| large | -74.8 |

Debt to GDP

| | |
|---|---|
| micro | -4.4 |
| tiny | 4.7 |
| small | -10.8 |
| medium | 3.1 |
| large | 16.6 |

Note: blue bar: significant at 10%; yellow bar: insignificant at 10%

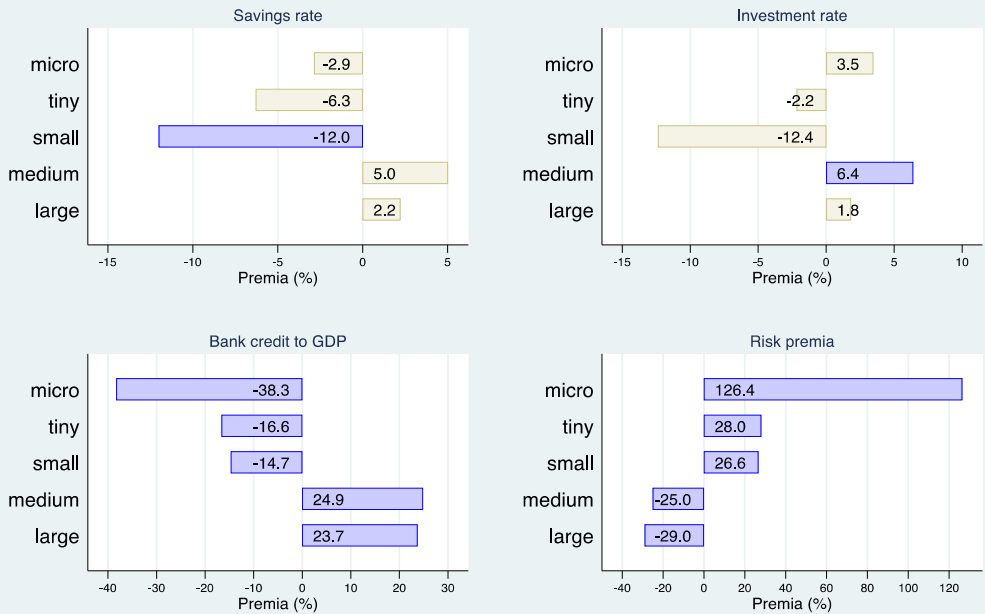

**Figure 3d: Size premia (%)**

Premia over group of small or large countries (by five country size groups)

Savings rate

| | |
|---|---|
| micro | -2.9 |
| tiny | -6.3 |
| small | -12.0 |
| medium | 5.0 |
| large | 2.2 |

Investment rate

| | |
|---|---|
| micro | 3.5 |
| tiny | -2.2 |
| small | -12.4 |
| medium | 6.4 |
| large | 1.8 |

Bank credit to GDP

| | |
|---|---|
| micro | -38.3 |
| tiny | -16.6 |
| small | -14.7 |
| medium | 24.9 |
| large | 23.7 |

Risk premia

| | |
|---|---|
| micro | 126.4 |
| tiny | 28.0 |
| small | 26.6 |
| medium | -25.0 |
| large | -29.0 |

Note: blue bar: significant at 10%; yellow bar: insignificant at 10%

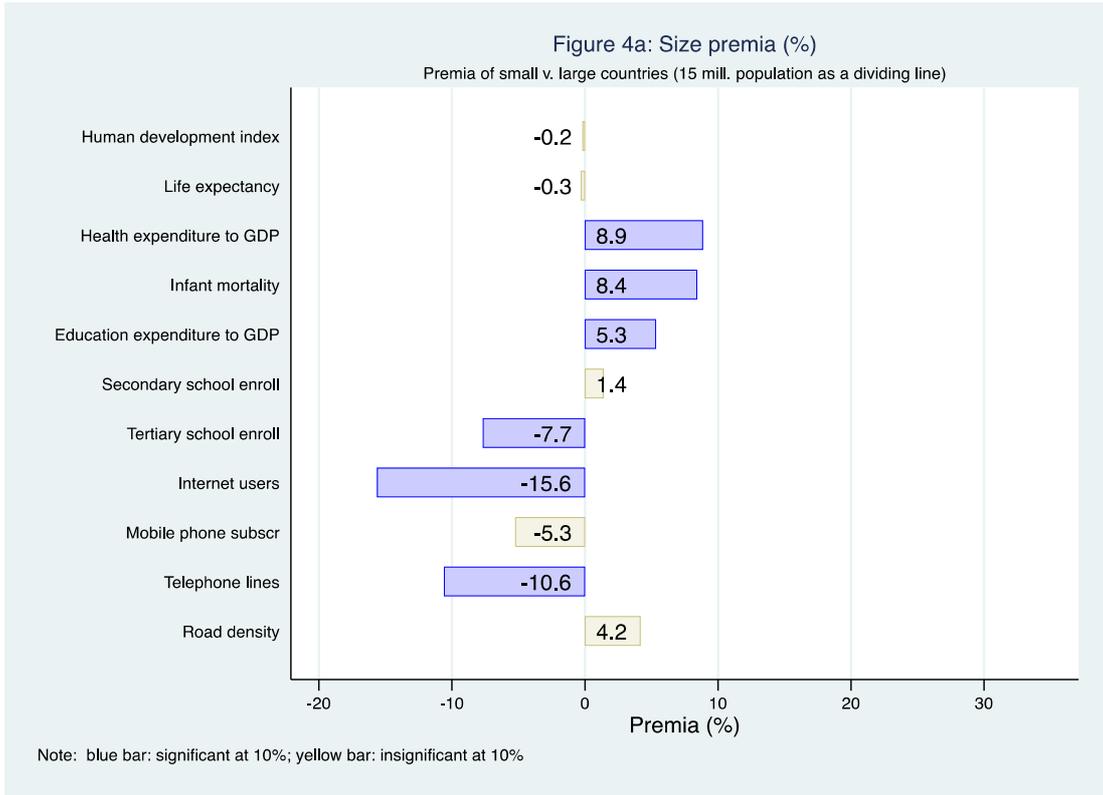

Figure 4a: Size premia (%)

Premia of small v. large countries (15 mill. population as a dividing line)

| Indicator | Value |
|---|---|
| Human development index | -0.2 |
| Life expectancy | -0.3 |
| Health expenditure to GDP | 8.9 |
| Infant mortality | 8.4 |
| Education expenditure to GDP | 5.3 |
| Secondary school enroll | 1.4 |
| Tertiary school enroll | -7.7 |
| Internet users | -15.6 |
| Mobile phone subscr | -5.3 |
| Telephone lines | -10.6 |
| Road density | 4.2 |

Note: blue bar: significant at 10%; yellow bar: insignificant at 10%

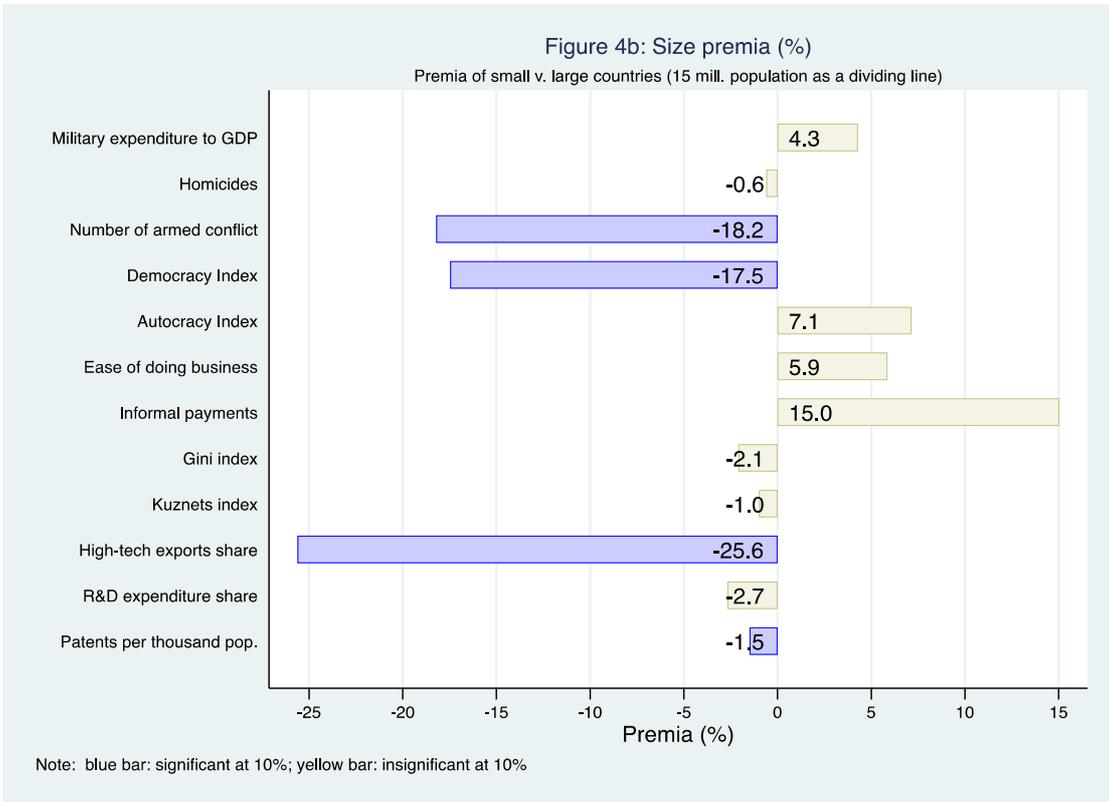

Figure 4b: Size premia (%)

Premia of small v. large countries (15 mill. population as a dividing line)

| Indicator | Value |
|---|---|
| Military expenditure to GDP | 4.3 |
| Homicides | -0.6 |
| Number of armed conflict | -18.2 |
| Democracy Index | -17.5 |
| Autocracy Index | 7.1 |
| Ease of doing business | 5.9 |
| Informal payments | 15.0 |
| Gini index | -2.1 |
| Kuznets index | -1.0 |
| High-tech exports share | -25.6 |
| R&D expenditure share | -2.7 |
| Patents per thousand pop. | -1.5 |

Note: blue bar: significant at 10%; yellow bar: insignificant at 10%

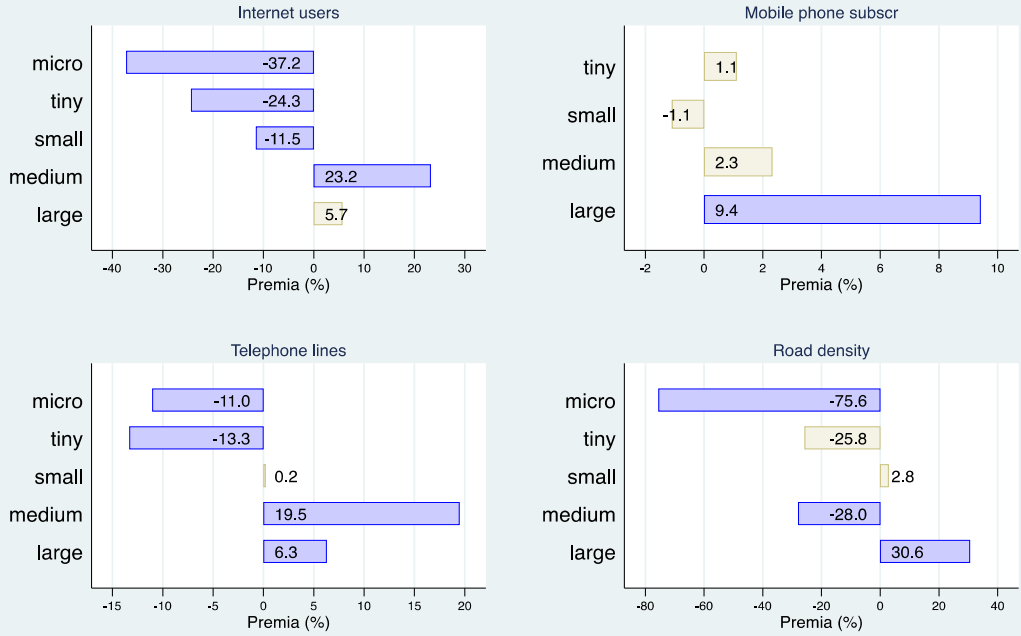

Figure 5a: Size premia (%)

Premia over group of small or large countries (by five country size groups)

Note: blue bar: significant at 10%; yellow bar: insignificant at 10%

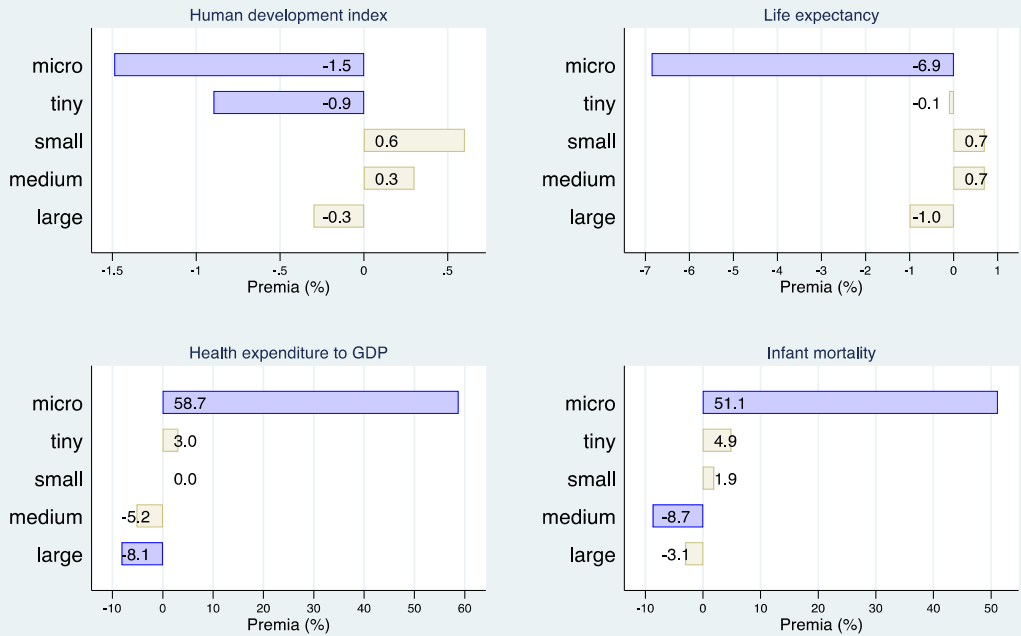

Figure 5b: Size premia (%)

Premia over group of small or large countries (by five country size groups)

Note: blue bar: significant at 10%; yellow bar: insignificant at 10%

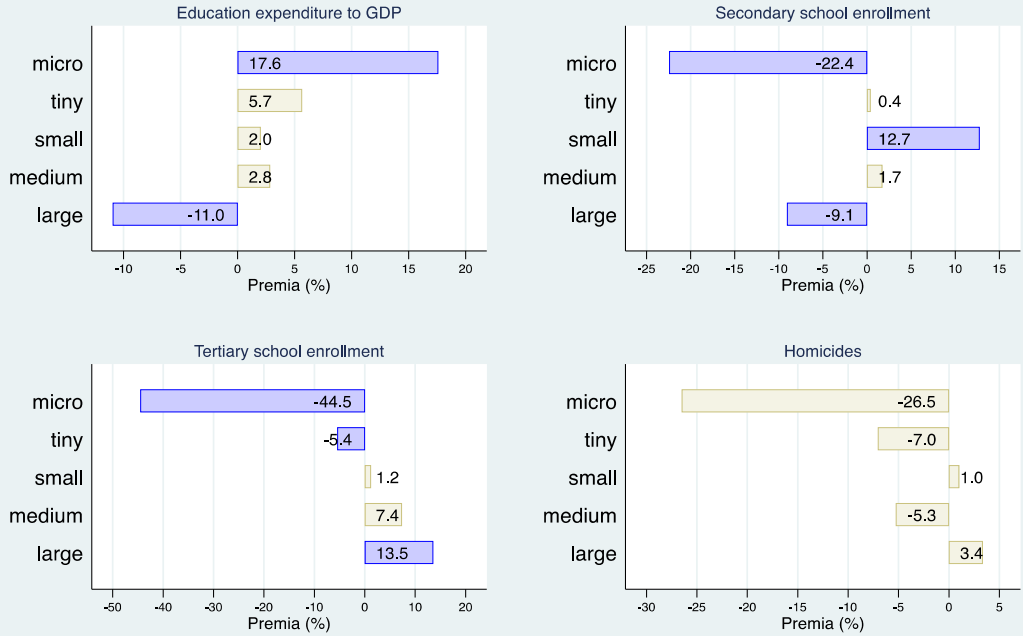

Figure 5c: Size premia (%)

Premia over group of small or large countries (by five country size groups)

Note: blue bar: significant at 10%; yellow bar: insignificant at 10%

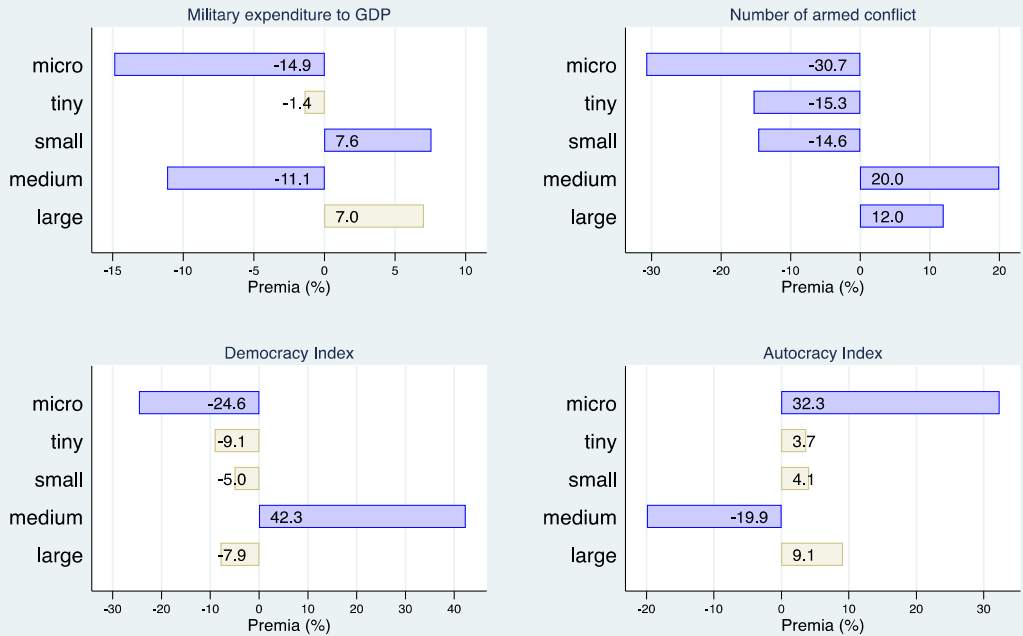

Figure 5d: Size premia (%)

Premia over group of small or large countries (by five country size groups)

Note: blue bar: significant at 10%; yellow bar: insignificant at 10%

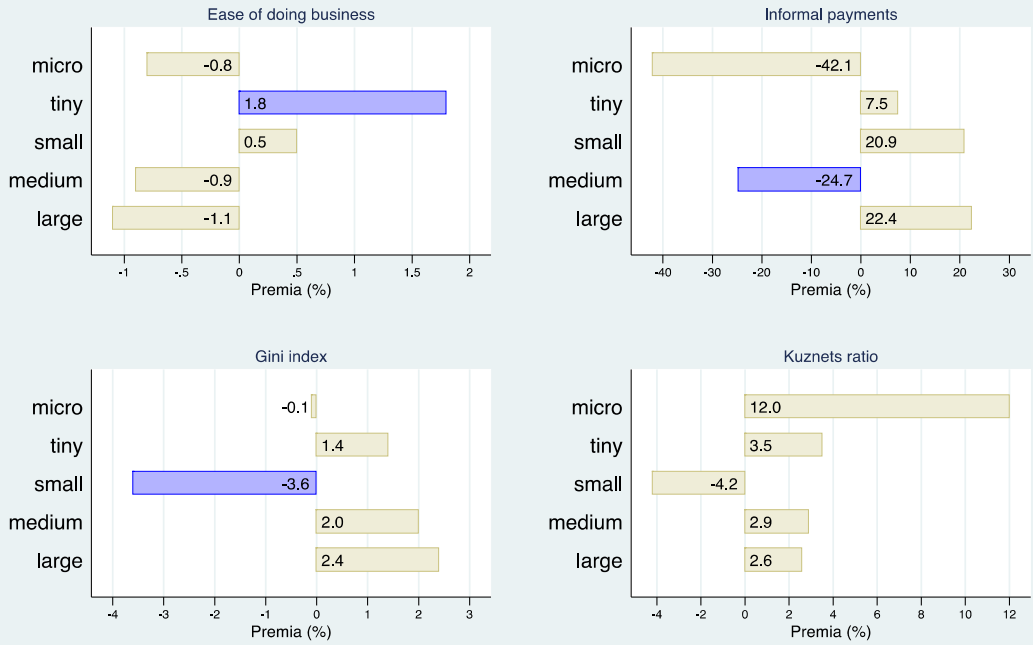

**Figure 5e: Size premia (%)**

Premia over group of small or large countries (by five country size groups)

Note: blue bar: significant at 10%; yellow bar: insignificant at 10%

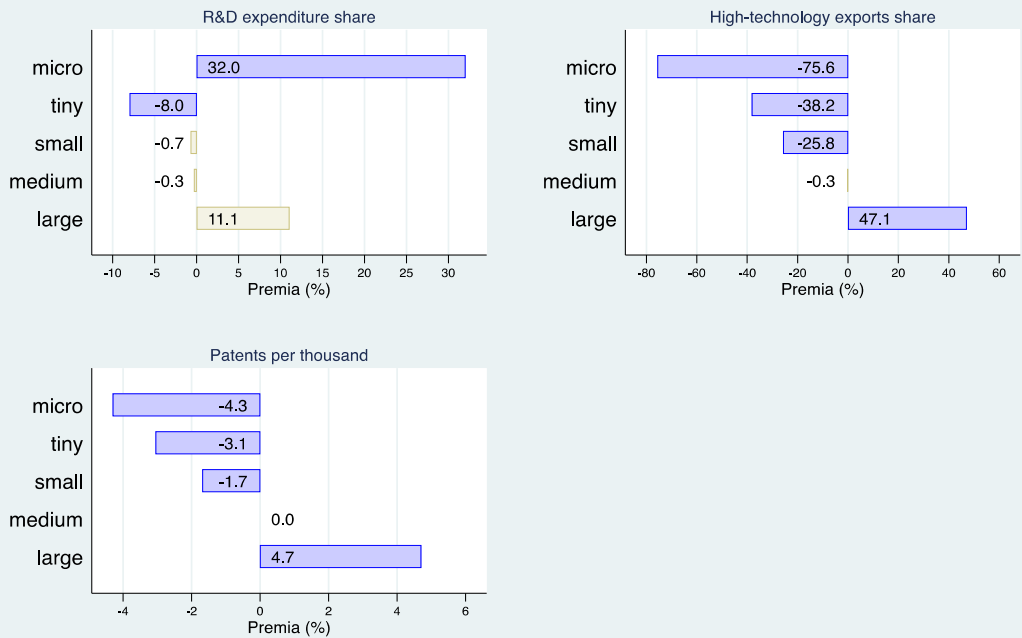

**Figure 5f: Size premia (%)**

Premia over group of small or large countries (by five country size groups)

Note: blue bar: significant at 10%; yellow bar: insignificant at 10%

# Appendix

Table A1: List of countries included in analysis with available data, in 1960 and 2015

| no. | 1960 | 2015 | no. | 2015 (cont.) |
|-----|------|------|-----|--------------|
| 1 | Argentina | Aruba | 97 | Cambodia |
| 2 | Australia | Andorra | 98 | Kiribati |
| 3 | Austria | Afghanistan | 99 | St. Kitts and Nevis |
| 4 | Burundi | Angola | 100 | Korea, Rep. |
| 5 | Belgium | Albania | 101 | Kosovo |
| 6 | Benin | United Arab Emirates | 102 | Kuwait |
| 7 | Burkina Faso | Argentina | 103 | Lao PDR |
| 8 | Bangladesh | Armenia | 104 | Lebanon |
| 9 | Bahamas, The | Antigua and Barbuda | 105 | Liberia |
| 10 | Belize | Australia | 106 | Libya |
| 11 | Bermuda | Austria | 107 | St. Lucia |
| 12 | Bolivia | Azerbaijan | 108 | Liechtenstein |
| 13 | Brazil | Burundi | 109 | Sri Lanka |
| 14 | Barbados | Belgium | 110 | Lesotho |
| 15 | Botswana | Benin | 111 | Lithuania |
| 16 | Central African Republic | Burkina Faso | 112 | Luxembourg |
| 17 | Canada | Bangladesh | 113 | Latvia |
| 18 | Chile | Bulgaria | 114 | Macao SAR, China |
| 19 | China | Bahrain | 115 | Morocco |
| 20 | Cote d'Ivoire | Bahamas, The | 116 | Monaco |
| 21 | Cameroon | Bosnia and Herzegovina | 117 | Moldova |
| 22 | Congo, Rep. | Belarus | 118 | Madagascar |
| 23 | Colombia | Belize | 119 | Maldives |
| 24 | Costa Rica | Bermuda | 120 | Mexico |
| 25 | Denmark | Bolivia | 121 | Marshall Islands |
| 26 | Dominican Republic | Brazil | 122 | Macedonia, FYR |
| 27 | Algeria | Barbados | 123 | Mali |
| 28 | Ecuador | Brunei Darussalam | 124 | Malta |
| 29 | Egypt, Arab Rep. | Bhutan | 125 | Montenegro |
| 30 | Spain | Botswana | 126 | Mongolia |
| 31 | Finland | Central African | 127 | Mozambique |
| 32 | Fiji | Canada | 128 | Mauritania |
| 33 | France | Switzerland | 129 | Mauritius |
| 34 | Gabon | Channel Islands | 130 | Malawi |
| 35 | United Kingdom | Chile | 131 | Malaysia |
| 36 | Ghana | China | 132 | Namibia |
| 37 | Greece | Cote d'Ivoire | 133 | Niger |
| 38 | Guatemala | Cameroon | 134 | Nigeria |
| 39 | Guyana | Congo, Rep. | 135 | Nicaragua |
| 40 | Honduras | Colombia | 136 | Netherlands |
| 41 | Hungary | Comoros | 137 | Norway |
| 42 | Indonesia | Cape Verde | 138 | Nepal |
| 43 | India | Costa Rica | 139 | New Zealand |
| 44 | Iceland | Cuba | 140 | Oman |
| 45 | Israel | Cyprus | 141 | Pakistan |
| 46 | Italy | Czech Republic | 142 | Panama |
| 47 | Japan | Germany | 143 | Peru |
| 48 | Kenya | Djibouti | 144 | Philippines |
| 49 | Korea, Rep. | Dominica | 145 | Palau |
| 50 | Liberia | Denmark | 146 | Papua New Guinea |
| 51 | Sri Lanka | Dominican Republic | 147 | Poland |
| 52 | Lesotho | Algeria | 148 | Puerto Rico |
| 53 | Luxembourg | Ecuador | 149 | Portugal |
| 54 | Morocco | Egypt, Arab Rep. | 150 | Paraguay |
| 55 | Madagascar | Eritrea | 151 | Qatar |
| 56 | Mexico | Spain | 152 | Romania |
| 57 | Mauritania | Estonia | 153 | Russian Federation |
| 58 | Malawi | Ethiopia | 154 | Rwanda |
| 59 | Malaysia | Finland | 155 | Saudi Arabia |
| 60 | Niger | Fiji | 156 | Sudan |
| 61 | Nigeria | France | 157 | Senegal |
| 62 | Nicaragua | Faeroe Islands | 158 | Singapore |
| 63 | Netherlands | Micronesia, Fed. Sts. | 159 | Solomon Islands |
| 64 | Norway | Gabon | 160 | Sierra Leone |
| 65 | Nepal | United Kingdom | 161 | El Salvador |
| 66 | Oman | Georgia | 162 | San Marino |
| 67 | Pakistan | Ghana | 163 | Serbia |
| 68 | Panama | Guinea | 164 | Sao Tome and Principe |
| 69 | Peru | Gambia, The | 165 | Suriname |
| 70 | Philippines | Guinea-Bissau | 166 | Slovak Republic |
| 71 | Papua New Guinea | Equatorial Guinea | 167 | Slovenia |
| 72 | Puerto Rico | Greece | 168 | Sweden |
| 73 | Portugal | Grenada | 169 | Swaziland |
| 74 | Paraguay | Greenland | 170 | Seychelles |
| 75 | Rwanda | Guatemala | 171 | Syrian Arab Republic |
| 76 | Sudan | Guyana | 172 | Chad |
| 77 | Senegal | Hong Kong SAR, China | 173 | Togo |
| 78 | Singapore | Honduras | 174 | Thailand |
| 79 | Sierra Leone | Croatia | 175 | Tajikistan |
| 80 | El Salvador | Haiti | 176 | Turkmenistan |
| 81 | Sweden | Hungary | 177 | Timor-Leste |
| 82 | Seychelles | Indonesia | 178 | Tonga |
| 83 | Syrian Arab Republic | Isle of Man | 179 | Trinidad and Tobago |
| 84 | Chad | India | 180 | Tunisia |
| 85 | Togo | Ireland | 181 | Turkey |
| 86 | Thailand | Iran, Islamic Rep. | 182 | Tuvalu |
| 87 | Trinidad and Tobago | Iraq | 183 | Tanzania |
| 88 | Turkey | Iceland | 184 | Uganda |
| 89 | Uruguay | Israel | 185 | Ukraine |
| 90 | United States | Italy | 186 | Uruguay |
| 91 | Vincent and the | Jamaica | 187 | United States |
| 92 | Venezuela, RB | Jordan | 188 | Uzbekistan |
| 93 | South Africa | Japan | 189 | Vincent and the |
| 94 | Congo, Dem. Rep. | Kazakhstan | 190 | Venezuela, RB |
| 95 | Zambia | Kenya | 191 | Vietnam |
| 96 | Zimbabwe | Kyrgyz Republic | 192 | Vanuatu |
| | | | 193 | West Bank and Gaza |
| | | | 194 | Samoa |
| | | | 195 | Yemen, Rep. |
| | | | 196 | South Africa |
| | | | 197 | Congo, Dem. Rep. |
| | | | 198 | Zambia |
| | | | 199 | Zimbabwe |

| Variable | (1)<br>log (Pop)<br>OLS | (2)<br>log (Pop)<br>FE | (3)<br>Size 10<br>OLS | (4)<br>Size 15<br>OLS | (5)<br>Size 20<br>OLS | C |
|---|---|---|---|---|---|---|
| GDP per capita (constant 2005 US$) | -0.257<br>[-21.73]*** | -0.546<br>[-11.16]*** | 0.376<br>[8.10]*** | 0.459<br>[9.95]*** | 0.432<br>[9.50]*** | 1, |
| Avg. annual GDP growth over past 5 years | -0.242<br>[-2.66]*** | 0.808<br>[1.14] | 0.189<br>[0.75] | 0.014<br>[0.06] | 0.073<br>[0.31] | 1, |
| Unemployment, total (% of total labor force) | -0.002<br>[-0.11] | -0.552<br>[-3.25]*** | 0.008<br>[0.14] | -0.004<br>[-0.06] | -0.029<br>[-0.60] | 6 |
| Central gov budget revenues (% of GDP) | -0.080<br>[-8.00]*** | 0.467<br>[3.44]*** | 0.192<br>[6.80]*** | 0.146<br>[3.82]*** | 0.113<br>[2.95]*** | 7 |
| Central gov budget expenditures (% of GDP) | -0.060<br>[-6.57]*** | -0.082<br>[-0.72] | 0.110<br>[4.62]*** | 0.081<br>[2.74]*** | 0.061<br>[2.17]** | 7 |
| Budget surplus/deficit (% of GDP) | -0.220<br>[-2.26]** | 0.253<br>[0.18] | 0.587<br>[2.34]** | 1.000<br>[4.07]*** | 0.928<br>[3.04]*** | 2 |
| Central government debt, total (% of GDP) | -0.000<br>[-0.01] | -2.212<br>[-4.90]*** | 0.016<br>[0.23] | 0.010<br>[0.14] | -0.041<br>[-0.54] | 6 |
| Inflation, consumer prices (annual %) | 0.065<br>[2.81]*** | 1.285<br>[6.74]*** | -0.171<br>[-2.34]** | -0.207<br>[-2.71]*** | -0.230<br>[-2.94]*** | 1, |
| Gross domestic savings (% of GDP) | 0.029<br>[1.73]* | 0.585<br>[5.21]*** | -0.080<br>[-1.65]* | -0.106<br>[-2.36]** | -0.102<br>[-2.38]** | 1, |
| Gross fixed capital formation (% of GDP) | -0.021<br>[-2.54]** | 0.495<br>[8.77]*** | 0.008<br>[0.33] | -0.051<br>[-2.20]** | -0.055<br>[-2.50]** | 1, |
| Country risk | 0.007<br>[0.93] | | -0.028<br>[-1.25] | -0.013<br>[-0.53] | -0.019<br>[-0.80] | 1 |
| Risk premium on lending ( %) | -0.037<br>[-1.20] | 0.114<br>[0.38] | 0.268<br>[2.41]** | 0.293<br>[2.45]** | 0.188<br>[1.71]* | 5 |
| Domestic banking credit (% of GDP) | 0.102<br>[6.04]*** | -0.166<br>[-0.86] | -0.229<br>[-4.92]*** | -0.282<br>[-5.47]*** | -0.253<br>[-5.26]*** | 1, |
| Trade (% of GDP) | -0.152<br>[-17.84]*** | -0.006<br>[-0.14] | 0.366<br>[14.28]*** | 0.458<br>[15.90]*** | 0.452<br>[15.67]*** | 1, |
| Current account balance (% of GDP) | 0.870<br>[3.49]*** | 4.503<br>[1.92]* | -1.339<br>[-2.15]** | -1.131<br>[-1.62] | -1.259<br>[-2.29]** | 1, |
| Democracy Index | -0.353<br>[-2.27]** | 12.082<br>[3.83]*** | 1.845<br>[3.88]*** | 1.432<br>[2.58]** | 1.015<br>[2.32]** | 7 |
| Public spending on education, total (% of GDP) | -0.021<br>[-1.83]* | -0.020<br>[-0.25] | 0.021<br>[0.60] | 0.070<br>[1.98]** | 0.074<br>[2.16]** | 6 |
| School enrollment, secondary (% gross) | 0.004<br>[0.36] | 0.992<br>[9.79]*** | -0.066<br>[-2.00]** | -0.013<br>[-0.38] | 0.004<br>[0.12] | 9 |
| School enrollment, tertiary (% gross) | 0.093<br>[5.73]*** | 0.085<br>[0.83] | -0.146<br>[-3.43]*** | -0.118<br>[-2.71]*** | -0.096<br>[-2.27]** | 8 |
| Human Development Index | -0.000<br>[-0.26] | 0.054<br>[5.73]*** | -0.007<br>[-1.47] | -0.005<br>[-1.01] | -0.005<br>[-1.01] | 3 |
| Life expectancy at birth, total (years) | 0.003<br>[1.92]* | 0.121<br>[10.74]*** | -0.021<br>[-3.87]*** | -0.008<br>[-1.30] | 0.000<br>[0.05] | 1, |
| Health expenditure, public (% of GDP) | -0.054<br>[-4.59]*** | 0.039<br>[0.28] | 0.063<br>[2.00]** | 0.090<br>[2.70]*** | 0.041<br>[1.22] | 7 |
| Income share held by highest 20% | -0.028 | 0.241 | 0.128 | 0.084 | -0.012 | 1, |

| | (1) | (2) | (3) | (4) | (5) | |
|---|---|---|---|---|---|---|
| | [-3.83]*** | [4.47]*** | [5.10]*** | [3.24]*** | [-0.49] | |
| Internet users (per 100 people) | 0.029 | -1.802 | -0.095 | -0.177 | -0.125 | 1, |
| | [1.87]* | [-5.62]*** | [-1.83]* | [-3.16]*** | [-2.06]** | |
| Mobile cellular subscriptions (per 100 people) | 0.043 | -1.021 | -0.030 | -0.063 | -0.058 | 1, |
| | [3.52]*** | [-12.55]*** | [-0.76] | [-1.52] | [-1.38] | |
| Telephone lines (per 100 people) | 0.015 | -0.162 | -0.070 | -0.138 | -0.096 | 1, |
| | [1.70]* | [-2.43]** | [-2.38]** | [-4.61]*** | [-3.09]*** | |
| Road density (km of road per 100 sq. km of land area) | 0.054 | 0.610 | -0.316 | -0.013 | -0.133 | 1 |
| | [0.87] | [1.70]* | [-2.00]** | [-0.06] | [-0.77] | |
| Military expenditure (% of GDP) | 0.014 | -0.039 | -0.001 | 0.046 | 0.015 | 6 |
| | [1.02] | [-0.28] | [-0.02] | [1.12] | [0.36] | |
| Intentional homicides (per 100,000 people) | 0.007 | 1.039 | 0.038 | -0.071 | -0.009 | 3 |
| | [0.24] | [2.74]*** | [0.51] | [-0.80] | [-0.09] | |
| No. of armed conflict | 0.093 | 0.002 | -0.112 | -0.181 | -0.190 | 5 |
| | [6.42]*** | [0.02] | [-2.56]** | [-4.23]*** | [-4.46]*** | |
| Log Km from equator | 0.046 | -0.272 | 0.078 | -0.101 | -0.039 | 5 |
| | [1.38] | [-1.17] | [0.95] | [-1.14] | [-0.44] | |
| Autocracy Index | -0.032 | 0.273 | -0.122 | 0.044 | 0.018 | 5 |
| | [-0.99] | [1.25] | [-1.56] | [0.51] | [0.21] | |
| Standard deviation of GDP growth over past 5 years | -0.080 | -0.560 | 0.197 | 0.200 | 0.159 | 1, |
| | [-6.68]*** | [-5.36]*** | [5.08]*** | [4.88]*** | [3.76]*** | |
| Ease of doing business index | -0.045 | | -0.050 | -0.046 | -0.118 | 1 |
| | [-1.28] | | [-0.44] | [-0.35] | [-0.83] | |
| Informal payments to public officials (% of firms) | 0.054 | -0.879 | 0.321 | 0.121 | 0.213 | |
| | [0.50] | [.] | [1.40] | [0.34] | [0.61] | |
| Foreign direct investment, net inflows (% of GDP) | -0.125 | 0.552 | 0.318 | 0.310 | 0.355 | 8 |
| | [-5.35]*** | [1.73]* | [4.35]*** | [3.75]*** | [4.75]*** | |
| Gini Index | -0.008 | 0.114 | -0.005 | 0.015 | 0.028 | 1 |
| | [-0.67] | [0.27] | [-0.17] | [0.46] | [0.91] | |
| Income share held by highest 10% | -0.001 | 0.143 | -0.005 | 0.031 | 0.033 | 1 |
| | [-0.08] | [0.38] | [-0.16] | [1.00] | [1.10] | |
| Kuznets ratio (highest 20 to lowest 40 % of income) | -0.030 | 0.164 | 0.024 | 0.060 | 0.077 | 1 |
| | [-1.08] | [0.23] | [0.39] | [0.93] | [1.24] | |
| Poverty gap at national poverty line (%) | -0.083 | -2.061 | 0.277 | 0.449 | 0.129 | |
| | [-1.16] | [-0.28] | [1.12] | [1.74]* | [0.71] | |
| High technology exports share (% manuf. exports) | 0.143 | 1.415 | -0.374 | -0.340 | -0.172 | 2 |
| | [2.98]*** | [0.77] | [-2.63]*** | [-2.28]** | [-1.17] | |
| Patents applied by residents (per million people) | 0.038 | -0.004 | -0.055 | -0.093 | -0.094 | 6 |
| | [4.22]*** | [-0.12] | [-2.92]*** | [-4.03]*** | [-3.85]*** | |
| R&D expenditure share (% GDP) | 0.054 | -0.197 | -0.049 | -0.035 | -0.080 | 3 |
| | [3.90]*** | [-1.66]* | [-1.33] | [-1.02] | [-2.24]** | |

*Notes*: Results of estimating model (1). Coefficients of regressions of log GDP per capita (2005 constant $) on indicated size indicator. Each row represents a separate regression. Model is estimated for the pooled sample and includes complete set of control variables (Controls 2) as explained in Section 3.1.

Robust t-statistics in brackets; *** p<0.01, ** p<0.05, * p<0.1.

Table A3: Coefficients for premia calculated in Figures 3, 4a, and 4b

| Variable | Coef. | t-stat | Obs. | R-sq. |
|---|---|---|---|---|
| GDP per capita | 0.451 | [9.98]*** | 1,809 | 0.820 |
| GDP growth | -0.153 | [-0.81] | 1,600 | 0.196 |
| Std. deviation of GDP growth | 0.173 | [4.56]*** | 1,469 | 0.212 |
| Openness | 0.452 | [16.65]*** | 1,563 | 0.604 |
| Current account to GDP | 0.452 | [3.55]*** | 316 | 0.546 |
| Net FDI to GDP | 0.340 | [5.22]*** | 830 | 0.330 |
| Unemployment rate | -0.047 | [-1.08] | 559 | 0.496 |
| CPI | -0.112 | [-1.88]* | 1,268 | 0.395 |
| Savings rate | -0.051 | [-1.47] | 1,353 | 0.474 |
| Investment rate | -0.049 | [-2.36]** | 1,397 | 0.208 |
| Government revenues to GDP | 0.081 | [3.25]*** | 711 | 0.642 |
| Government expenditures to GDP | 0.048 | [2.17]** | 710 | 0.613 |
| Budget balance to GDP | 0.956 | [6.36]*** | 184 | 0.611 |
| Debt to GDP | -0.074 | [-1.19] | 576 | 0.381 |
| Bank credit to GDP | -0.207 | [-5.18]*** | 1,436 | 0.617 |
| Country risk | -0.006 | [-0.45] | 85 | 0.914 |
| Risk premia | 0.327 | [4.82]*** | 490 | 0.536 |
| Human develop. index | -0.002 | [-0.64] | 295 | 0.958 |
| Life expectancy | -0.003 | [-0.64] | 1,544 | 0.849 |
| Health expenditure to GDP | 0.085 | [2.95]*** | 669 | 0.583 |
| Infant mortality | 0.081 | [3.24]*** | 1,659 | 0.887 |
| Education expenditure to GDP | 0.052 | [1.67]* | 609 | 0.357 |
| Secondary school enroll | 0.014 | [0.50] | 947 | 0.826 |
| Tertiary school enroll | -0.080 | [-2.08]** | 828 | 0.905 |
| Internet users | -0.170 | [-3.10]*** | 1,020 | 0.871 |
| Mobile phone subscr | -0.054 | [-1.39] | 1,756 | 0.923 |
| Telephone lines | -0.112 | [-4.00]*** | 1,672 | 0.914 |
| Road density | 0.041 | [0.39] | 107 | 0.946 |
| Military expenditure to GDP | 0.042 | [1.17] | 624 | 0.441 |
| Homicides | -0.006 | [-0.09] | 310 | 0.766 |
| Number of armed conflict | -0.201 | [-4.97]*** | 468 | 0.372 |
| Democracy Index | -0.192 | [-2.45]** | 472 | 0.655 |
| Autocracy Index | 0.069 | [0.90] | 471 | 0.636 |
| Ease of doing business | 0.057 | [0.75] | 119 | 0.890 |
| Informal payments | 0.140 | [1.67] | 37 | 0.897 |
| Gini index | -0.021 | [-0.83] | 108 | 0.918 |
| Kuznets ratio | -0.010 | [-0.17] | 102 | 0.912 |
| High-tech exports share | -0.296 | [-2.68]*** | 182 | 0.739 |
| R&D expenditure share | -0.027 | [-0.98] | 253 | 0.871 |
| Patents per thousand pop. | -0.015 | [-2.42]** | 612 | 0.723 |

*Notes:* Results of estimating model (1). Coefficients on Size dummy variable taking the value of 1 for population size smaller than 15 million, and 0 otherwise. Each row represents a separate regression. Regressions include the complete set of control variables as explained in Section 3.1. Control variables and the constant term are omitted from the presentation for brevity.

Robust t-statistics in brackets; *** p<0.01, ** p<0.05, * p<0.1.

(Full results can be obtained from authors upon request.)

## Table A4: Coefficients for premia calculated in Figures 3a – 3d, and 5a – 5f

| Variable | micro Coef. | t-stat | tiny Coef. | t-stat | small Coef. | t-stat | medium Coef. | t-stat | large Coef. | t-stat |
|---|---|---|---|---|---|---|---|---|---|---|
| GDP per capita | 0.828 | [9.34]*** | 0.432 | [9.14]*** | 0.179 | [3.92]*** | -0.303 | [-5.37]*** | -0.682 | [-12.49]** |
| GDP growth | -0.568 | [-1.55] | -0.093 | [-0.41] | -0.308 | [-1.58] | 0.123 | [0.54] | 0.099 | [0.41] |
| Std. deviation of GDP growth | 0.39 | [5.99]*** | 0.18 | [3.85]*** | 0.05 | [1.22] | -0.195 | [-4.17]*** | -0.153 | [-3.22]*** |
| Openness | 0.587 | [9.72]*** | 0.452 | [13.63]*** | 0.349 | [13.23]*** | -0.377 | [-12.44]*** | -0.512 | [-15.03]*** |
| Current account to GDP | 0.243 | [0.96] | 0.353 | [1.70]* | 0.332 | [2.34]** | -0.222 | [-1.22] | -0.484 | [-3.52]*** |
| Net FDI to GDP | 0.136 | [1.22] | 0.538 | [8.37]*** | 0.177 | [2.54]** | -0.264 | [-3.47]*** | -0.46 | [-5.44]*** |
| Unemployment rate | -0.378 | [-3.28]*** | 0.004 | [0.07] | 0.031 | [0.66] | 0.157 | [2.85]*** | -0.028 | [-0.54] |
| CPI | 0.013 | [0.13] | -0.287 | [-4.40]*** | -0.056 | [-0.80] | 0.062 | [0.84] | 0.073 | [0.95] |
| Savings rate | -0.029 | [-0.40] | -0.065 | [-1.45] | -0.128 | [-3.61]*** | 0.049 | [1.17] | 0.022 | [0.50] |
| Investment rate | 0.034 | [0.77] | -0.022 | [-0.87] | -0.132 | [-6.13]*** | 0.062 | [2.48]** | 0.018 | [0.68] |
| Government revenues to GDP | 0.25 | [5.35]*** | 0.045 | [1.45] | -0.055 | [-1.70]* | -0.073 | [-2.39]** | -0.109 | [-3.56]*** |
| Government expenditures to GD | 0.165 | [3.71]*** | 0.012 | [0.49] | -0.052 | [-2.10]** | -0.017 | [-0.62] | -0.072 | [-2.51]** |
| Budget balance to GDP | 0.205 | [0.43] | 0.877 | [4.01]*** | 0.62 | [3.00]*** | -0.807 | [-4.75]*** | -1.378 | [-4.62]*** |
| Debt to GDP | -0.045 | [-0.46] | 0.046 | [0.64] | -0.114 | [-1.86]* | 0.031 | [0.39] | 0.154 | [2.07]** |
| Bank credit to GDP | -0.483 | [-6.29]*** | -0.182 | [-3.29]*** | -0.159 | [-3.32]*** | 0.222 | [4.27]*** | 0.213 | [4.47]*** |
| Country risk | -0.037 | [-1.16] | 0.027 | [1.46] | -0.003 | [-0.17] | 0.017 | [0.91] | -0.016 | [-1.01] |
| Risk premia | 0.817 | [9.72]*** | 0.247 | [2.58]** | 0.236 | [2.63]*** | -0.288 | [-3.38]*** | -0.342 | [-4.76]*** |
| Human develop. index | -0.015 | [-1.94]* | -0.009 | [-2.06]** | 0.006 | [1.36] | 0.003 | [0.66] | -0.003 | [-0.66] |
| Life expectancy | -0.071 | [-7.39]*** | -0.001 | [-0.09] | 0.007 | [1.11] | 0.007 | [1.08] | -0.01 | [-1.49] |
| Health expenditure to GDP | 0.462 | [8.96]*** | 0.03 | [1.03] | 0 | [-0.01] | -0.053 | [-1.58] | -0.085 | [-2.39]** |
| Infant mortality | 0.413 | [9.95]*** | 0.048 | [1.57] | 0.019 | [0.66] | -0.091 | [-2.91]*** | -0.031 | [-0.97] |
| Education expenditure to GDP | 0.162 | [2.47]** | 0.055 | [1.64] | 0.02 | [0.65] | 0.028 | [0.73] | -0.116 | [-3.28]*** |
| Secondary school enroll | -0.254 | [-4.78]*** | 0.004 | [0.12] | 0.12 | [4.18]*** | 0.017 | [0.53] | -0.095 | [-2.84]*** |
| Tertiary school enroll | -0.589 | [-7.04]*** | -0.056 | [-1.23] | 0.012 | [0.33] | 0.071 | [1.46] | 0.127 | [2.72]*** |
| Internet users | -0.466 | [-4.66]*** | -0.279 | [-4.42]*** | -0.122 | [-2.03]** | 0.209 | [3.35]*** | 0.055 | [0.74] |
| Mobile phone subscr | -0.267 | [-3.86]*** | 0.011 | [0.26] | -0.011 | [-0.26] | 0.023 | [0.52] | 0.09 | [1.67]* |
| Telephone lines | -0.117 | [-2.36]** | -0.143 | [-4.17]*** | 0.002 | [0.06] | 0.178 | [5.54]*** | 0.061 | [1.65]* |
| Road density | -1.409 | [-4.92]*** | -0.298 | [-1.15] | 0.028 | [0.31] | -0.328 | [-2.37]** | 0.267 | [2.60]** |
| Military expenditure to GDP | -0.161 | [-2.17]** | -0.014 | [-0.33] | 0.073 | [1.90]* | -0.118 | [-2.97]*** | 0.068 | [1.42] |
| Homicides | -0.308 | [-1.39] | -0.073 | [-0.79] | 0.01 | [0.15] | -0.054 | [-0.52] | 0.033 | [0.46] |
| Number of armed conflict | -0.367 | [-2.56]** | -0.166 | [-2.97]*** | -0.158 | [-4.33]*** | 0.182 | [4.63]*** | 0.113 | [2.26]** |
| Democracy Index | -0.282 | [-2.04]** | -0.095 | [-1.14] | -0.051 | [-0.65] | 0.353 | [3.71]*** | -0.082 | [-0.92] |
| Autocracy Index | 0.28 | [2.04]** | 0.036 | [0.41] | 0.04 | [0.50] | -0.222 | [-2.58]** | 0.087 | [0.93] |
| Ease of doing business | 0.306 | [3.06]*** | -0.127 | [-1.18] | -0.077 | [-0.90] | -0.067 | [-0.77] | 0.089 | [0.75] |
| Informal payments | | | -1.187 | [-6.88]*** | | | -0.069 | [-0.43] | 0.337 | [1.70] |
| Gini index | 0.498 | [1.66] | -0.097 | [-2.17]** | -0.008 | [-0.28] | 0.008 | [0.24] | 0.038 | [0.76] |
| Kuznets ratio | | | -0.16 | [-1.59] | 0.059 | [0.74] | -0.015 | [-0.24] | 0.029 | [0.28] |
| High-tech exports share | -1.412 | [-4.85]*** | -0.481 | [-4.16]*** | -0.298 | [-2.65]*** | -0.003 | [-0.02] | 0.386 | [3.29]*** |
| R&D expenditure share | 0.278 | [3.62]*** | -0.083 | [-2.25]** | -0.007 | [-0.30] | -0.003 | [-0.11] | 0.105 | [3.24]*** |
| Patents per thousand pop. | -0.044 | [-1.89]* | -0.031 | [-3.53]*** | -0.017 | [-2.79]*** | 0 | [-0.01] | 0.046 | [4.22]*** |

*Notes*: Results of estimating model (1). Coefficients on five Size dummy variables. Each coefficient corresponds to a separate regression. Regressions include the complete set of control variables as explained in Section 3.1. Control variables and the constant term are omitted from the presentation for brevity.

Robust t-statistics in brackets; *** p<0.01, ** p<0.05, * p<0.1.

(Full results can be obtained from authors upon request.)